\documentclass[12pt]{iopart}

\usepackage{iopams}  
\usepackage{graphicx}
\usepackage{dcolumn}
\usepackage{bm}
\usepackage{cite}
\begin{document}

\title{Filling and wetting transitions on sinusoidal substrates: a mean-field study of the Landau-Ginzburg model.}

\author{\'Alvaro Rodr\'{\i}guez-Rivas}
\address{Departamento de F\'{\i}sica At\'omica, Molecular y
Nuclear, Area de F\'{\i}sica Te\'orica, Universidad de Sevilla,
Apartado de Correos 1065, 41080 Sevilla, Spain}
\author{Jos\'e Antonio Galv\'an Moreno}
\address{Departamento de F\'{\i}sica Aplicada II, Universidad de Sevilla,
ETS Arquitectura, Avenida Reina Mercedes 2, 41012 Sevilla, Spain}
\author{Jos\'e M. Romero-Enrique}
\address{Departamento de F\'{\i}sica At\'omica, Molecular y
Nuclear, Area de F\'{\i}sica Te\'orica, Universidad de Sevilla,
Apartado de Correos 1065, 41080 Sevilla, Spain}
\begin{abstract}
We study the interfacial phenomenology of a fluid in contact with a microstructured substrate within the mean-field 
approximation. The sculpted substrate is a one-dimensional array of infinitely long grooves of sinusoidal
section of periodicity length $L$ and amplitude $A$. The system is modelled using the Landau-Ginzburg functional, with 
fluid-substrate couplings which 
correspond to either first-order or critical wetting for a flat substrate. We investigate the effect of the roughness 
of the substrate in the interfacial phenomenology, paying special attention to filling and wetting phenomena, 
and compare the results with the predictions of the macroscopic and interfacial Hamiltonian theories. At bulk 
coexistence, for values of $L$ much larger than the bulk correlation, we observe first-order filling transitions 
between dry and partially filled interfacial states, which extend off-coexistence, ending at a critical point; and wetting 
transitions between partially filled and completely wet interfacial states with the same order as for the flat substrate
(if first-order, wetting extends off-coexistence in a prewetting line). 
On the other hand, if the groove height is of order of the correlation length, only wetting transitions between
dry and complete wet states are observed. However, their characteristics depend on the order of the wetting transition
for the flat substrate. So, if it is first-order, the wetting transition temperature for the rough substrate is reduced 
with respect to the wetting transition temperature for a flat substrate, and coincides with the
Wenzel law prediction for very shallow substrates. On the contrary, if the flat substrate wetting transition is 
continuous, the roughness does not change the wetting temperature. The filling transition for shallow substrates disappears
at a triple point for first-order wetting substrates, and at a critical point for critical wetting substrates. 
The macroscopic theory only describes accurately the filling transition close to bulk coexistence and large $L$, while
microscopic structure of the fluid is essential to understand wetting and filling away from bulk coexistence. 
\end{abstract}

\maketitle

\section{Introduction\label{sec1}}

The interfacial behaviour of fluids in contact with structured substrates has received considerable attention in the last
years \cite{Quere}. This research is essential for the lab-on-a-chip applications, which aim to miniaturize chemical
plants to chip format \cite{Service,Herminghaus}. The advances in lithography techniques allow to sculpt grooves of 
controlled geometry on solid substrates on the micro- and nano-scales \cite{Whitesides}, which has allowed experimental 
studies of the influence of the geometry on fluid adsorption \cite{Bruschi,Bruschi2,Bruschi3,Bruschi4,Gang,Hofmann,Checco,
Javadi}. From a theoretical point of view, wetting and related phenomena at planar substrates have been studied in depth for
simple fluids \cite{Margarida,Dietrich,Schick,Forgacs}. For rough substrates, macroscopic models lead to phenomenological
laws for fluid adsorption on rough substrates, such as the Wenzel law \cite{Wenzel,Wenzel2} and the Cassie-Baxter law \cite{Cassie}.
The adsorption of fluids on microstructured substrates shows distinct characteristics compared to planar systems 
\cite{Gau,Rascon2,Quere2}. An example of this feature is the filling transition observed in linear-wedge shaped grooves 
\cite{Concus,Pomeau,Hauge2}, in which the wedge is completely filled by liquid when the contact angle $\theta$ associated to the 
sessile droplet on a flat substrate is equal to the tilt angle $\alpha$ of the wedge with respect to the horizontal plane.
However, macroscopic arguments show that it may appear in other microstructured substrates \cite{Rejmer}.
The filling transition in the wedge has been studied extensively in last years \cite{Rejmer0,Parry,Parry2,Parry3,Bednorz,
Parry4,Abraham,Abraham2,Albano,Milchev,Milchev2,Greenall,JM,Binder,Henderson,Henderson2,Henderson3,Rascon3,JM1,JM2,JM3,
Parry5,Nelson,JM4,Malijevsky,Malijevsky2}. However, other substrate geometries have been also studied, such as capped 
capillaries \cite{Marconi,Darbellay,Parry6,Roth,Malijevsky0,Rascon4,Yatsyshin}, crenellated substrates 
\cite{Tasinkevych,Tasinkevych2,Malijevsky3}, parabolic pits \cite{Kubalski,Tasinkevych,Tasinkevych2,Checco}, 
and sinusoidal substrates \cite{Swain,Rascon,Rejmer2}. 
Earlier studies relied on interfacial Hamiltonian theories, but in recent times, density-functional theories have
been applied, from simple square-gradient functionals, to more sofisticated functionals such as the fundamental measure 
theory. 

In this paper we revisit the study of fluid adsorption on sinusoidal substrates, paying special attention to filling and 
wetting transitions. Unlike previous interfacial Hamiltonian studies \cite{Rascon,Rejmer2}, which were restricted to shallow 
substrates, we consider intermediate and large values of the roughness. We consider a more microscopic model 
(the Landau-Ginzburg model), which allows us to extract the order parameter profile (i.e. density in our case) by 
minimizing a square-gradient functional. From these results, we are able to locate the gas-liquid interface (by imposing
a crossing criterion on the order parameter profile) and the different interfacial transitions are obtained by the 
crossing or merging of the free-energy branches associated to different interfacial states as the thermodynamic fields are 
changed. The adsorption phenomenology will be analyzed and compared with previous approaches, in particular the 
macroscopic theory and interfacial Hamiltonian model studies. 

The paper is organized as follows. Section \ref{sec2} is devoted to the description of the macroscopic theory of adsorption
and its application to sinusoidal substrates. In section \ref{sec3} we explain the numerical methodology, while 
the results are described in section \ref{sec4}. We present and discuss the main conclusions of our study in section 
\ref{sec5}. The paper concludes with an appendix where we revisit the wetting behaviour of the Landau-Ginzburg model
for a flat substrate. 

\section{Macroscopic theory of adsorption \label{sec2}}
\begin{figure}[t]
\centerline{\includegraphics[width=\columnwidth]{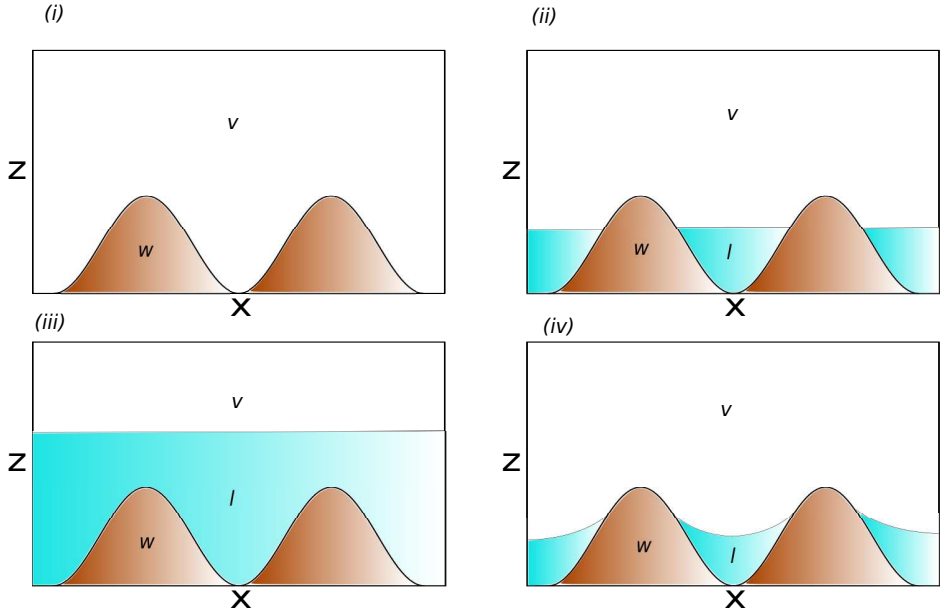}}
\caption
{Schematic representation of interfacial states. For bulk coexistence, (i) represents
 the dry (D) state, (ii) represents the partially filled (F) state and (iii) corresponds
to the completely wet state. In (iv) we represent the typical interfacial profile of the
partially filled state out of bulk coexistence. The substrate is
labelled by $w$, $l$ stands for the liquid and $v$ for the vapor.}
\label{fig1}
\end{figure}

From a macroscopic point of view we can understand much of the phenomenology of fluid adsorption on rough substrates.
Consider a gas at saturation conditions (i.e. coexisting with a liquid) in the presence of a rough substrate, 
which we will consider translationally invariant along the $y$-direction (with a total length $L_y$) and periodic 
across the $x$-direction, with a period $L$ much larger than the typical molecular lengthscales. 
We assume that the substrate favors nucleation of the liquid phase on its surface. The height of the substrate 
is given by a smooth function $\psi(x)$, which will be assumed to be an even function in its argument. 
There are three possible situations which the system may present 
\cite{Rejmer}: the interfacial dry state ($D$), in which only a thin (microscopic) liquid layer is adsorbed on 
the substrate; the partially filled state ($F$), in which the substrate grooves are partially filled with liquid 
up to a height $h=\psi(x_c)$; and the complete wet state ($W$), in which a thick liquid macroscopic layer between 
the substrate and the vapor is formed (see figure \ref{fig1}).

To see the relative stability of these phases, we consider the excess surface free energy ${\cal F}$ with respect to the 
bulk for each state. In the macroscopic approach, we assume that interactions between the different interfaces is negligible,
so the surface free energy can be obtained directly as the sum of the contributions of each surface/interface. We denote 
${\cal S}$ as the total area of the substrate and ${\cal A}$ as its projection on the plane $x-y$ plane.
The surface free energy of the $D$ state, ${\cal F}^{D}$, can be obtained from the substrate surface tension between 
\emph{flat} sustrate and a vapour in bulk, $\sigma_ {vw}$, as:
\begin{equation}
{\cal F}^{D}={\cal S}\sigma_{vw}
\label{free-energy-dry}
\end{equation}
On the other hand, the surface free energy of the partially filled state ${\cal F}^F$ is given by the expression:
\begin{equation}
{\cal F}^{F}=S(x_c)\sigma_{lw}+({\cal S}-S(x_c))\sigma_{vw}+2{\cal A} \frac{x_c}{L}
\sigma_{lv}
\label{free-energy-filled}
\end{equation}
where $\sigma_ {lw}$ is the interfacial tension between the liquid and the flat substrate, $\sigma_{lv}$ is the 
surface tension associated to the liquid-vapor interface, and $S(x_c)$ is the substrate area in contact with the 
liquid phase, which can be obtained from the value of $x=x_c$ at which the liquid-vapour interface is in contact with  
the substrate as
\begin{equation}
S(x_c)={\cal S}\frac{\int^{x_c}_{-{x_c}}\sqrt{1+\psi_x^2}dx}{\int^{L/2}_{-L/2}\sqrt{1+\psi_x^2}dx}
\end{equation}
where $\psi_x$ represents the derivative of $\psi$ with respect to $x$. Note that the free energy ${\cal F}^{D}$ given by 
(\ref{free-energy-dry}) corresponds to the limit $x_c \to 0$ from (\ref{free-energy-filled}).
The value of $x_c$ can be obtained from the minimization of free energy (\ref{free-energy-filled}) with respect to 
that parameter. By Young's law, which relates the different surface tensions with the surface contact angle $\theta$:
\begin{equation}
\sigma_{vw}-\sigma_{lw}=\sigma_{lv}\cos\theta
\label{Young}
\end{equation}
the free energy ${\cal F}^{F}$, (\ref{free-energy-filled}) can be rewritten as:
\begin{equation}
{\cal F}^{F}={\cal F}^{D}+\frac{2{\cal A}}{L}\sigma_{lv}\left(x_c-\cos\theta
\int_0^{x_c}\sqrt{1+\psi_x^ 2(x)}dx\right)
\label{free-energy-filled-2}
\end{equation}
Therefore, as the derivative of this function with respect to $x$ must vanish at $x=x_c$, the following condition is 
satisfied \cite{Rejmer}:
\begin{equation}
0=(1-\sqrt{1+\psi_x^ 2(x_c)}\cos\theta)=\left(1-\frac{\cos\theta}{\cos\alpha}\right)
\label{filling-equil-cond}
\end{equation}
where $\alpha$ is the angle between the liquid-vapor interface and the substrate at the contact $x=x_c$.
This result has a clear physical interpretation: the filled region by liquid should make contact with the substrate at 
the point where $\alpha$ is equal to the contact angle $\theta$. However, this solution is only a local minimum 
if $(d\psi(x_c)/dx)\times (d^2 \psi(x_c)/dx^2)<0$ \cite{Rejmer}.
Finally, the interfacial free energy for the state of complete wet state ${\cal F}^W$  is given by:
\begin{equation}
{\cal F}^{W}={\cal S}\sigma_{lw}+{\cal A}\sigma_{lv}
\label{free-energy-wet}
\end{equation}
Notice that macroscopically this expression corresponds to the limit $x_c \to L/2$ of (\ref{free-energy-filled}).

Several transitions between the different interfacial states may be observed. At low temperatures the most
stable state is the dry state, whereas at high temperatures (i.e. above the wetting temperature of the flat substrate) the 
preferred state corresponds to complete wetting. Therefore, for intermediate temperatures
must exist phase transitions between different interfacial states. For example, a wetting transition between
$D$ and $W$ can occur when both states have the same free energy:
\begin{equation}
{\cal F}^{W}-{\cal F}^{D}=0={\cal S}(\sigma_{lw}-\sigma_{vw})+{\cal A}\sigma_{lv}
\label{df31}
\end{equation}
Using Young's law (\ref{Young}), we obtain the following condition for the wetting transition:
\begin{equation}
\frac{\cal S}{\cal A}\cos\theta=r\cos\theta=1
\label{wetting-wenzel}
\end{equation}
where the roughness parameter is defined as $r={\cal S}/{\cal A}$. This is precisely the result obtained by Wenzel law
\cite{Wenzel, Wenzel2}: the contact angle of a liquid drop on a rough substrate, $\theta_r$, is related to the contact 
angle of a flat substrate $\theta$ via the expression $\cos \theta_r = r \cos \theta$. Therefore, as the wetting 
transition occurs when $\theta_r \to 0 $, we recover the expression (\ref{wetting-wenzel}).

\begin{figure}[t]
\centerline{\includegraphics[width=\columnwidth]{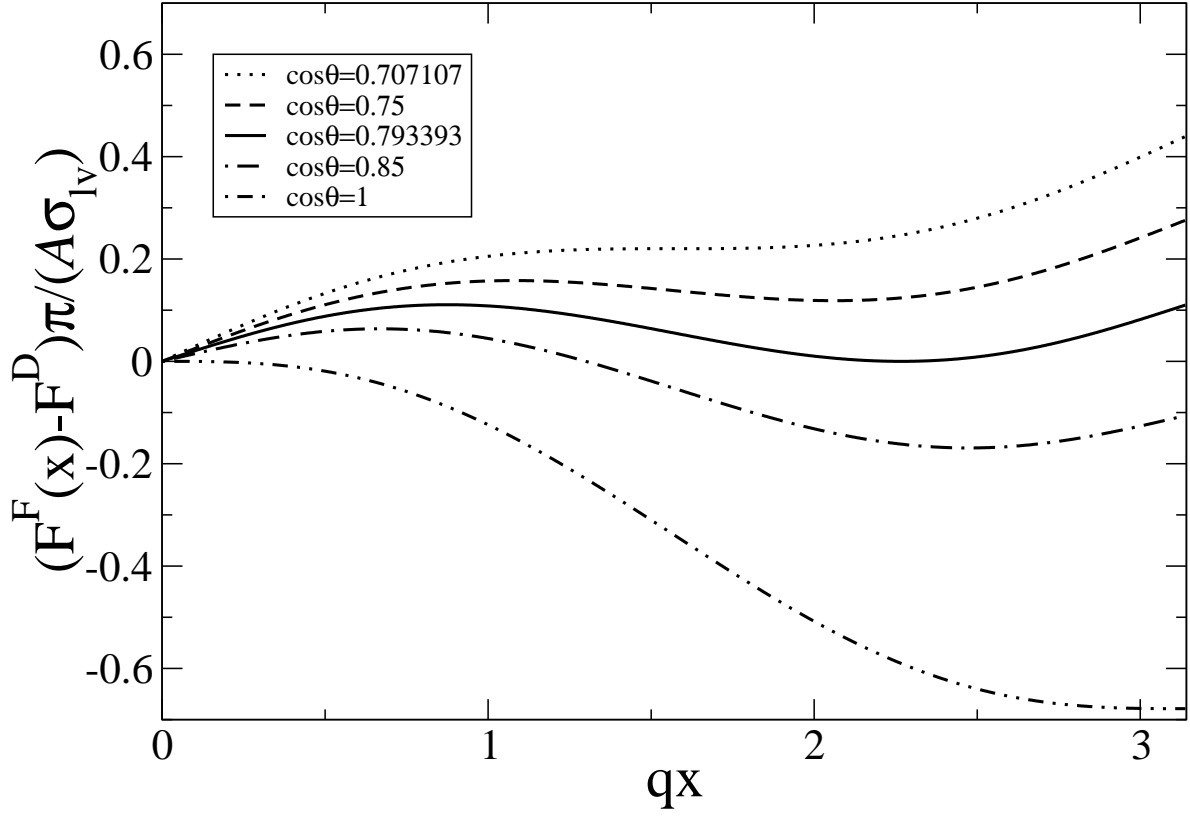}}
\caption
{Behaviour of ${\cal F}^{F}(x)$ as a function of $qx$ for $qA=1$ and contact angles:
$\theta=\theta^*$ (dotted line), $\theta_f<\theta<\theta^*$ (dashed line), $\theta=\theta_f$ (solid line),
$\theta<\theta_f$ (dot-dashed line), and $\theta=0$ (double dot-dashed line).
}
\label{fig2}
\end{figure}

It is also possible a transition from a dry state to a filled state. This transition is called in the literature
either \emph{filling} \cite{Rejmer} or \emph{unbending} \cite{Rascon} transition.
The filling transition occurs when:
\begin{eqnarray}
{\cal F}^{F}-{\cal F}^{D}=0\nonumber\\
=\frac{2{\cal A}}{L}\sigma_{lv}\left(x_c-\cos\theta
\int_0^{x_c}\sqrt{1+\psi_x^ 2(x)}dx\right)
\label{df21}
\end{eqnarray} 
which leads to the expression:
\begin{equation}
\frac{\int^{x_c}_{0}\sqrt{1+\psi_x^2}dx}{x_c}\cos\theta\equiv r_c\cos\theta=1
\end{equation}
where $r_c>1$. If this transition occurs at temperatures below the predicted by (\ref{wetting-wenzel}), then Wenzel 
law is no longer valid. In fact, under these conditions the macroscopic theory predicts that 
the wetting transition will occur between an $F$ and $W$ state when:
\begin{eqnarray}
{\cal F}^{W}-{\cal F}^{F}=0\nonumber\\
=({\cal S}-S(x_c))(\sigma_{lw}-\sigma_{vw})+{\cal A} \left(1-\frac{2x_c}{L}\right)\sigma_{lv}
\label{df23}
\end{eqnarray}
Now we will restrict ourselves to the sinusoidal substrate, characterized by an amplitude $A$ and a wavenumber $q=2\pi/L$, 
with the subtrate height $\psi(x)$ given by:
\begin{equation}
\psi(x)=A(1-\cos qx)
\label{defpsi}
\end{equation}
For this substrate, $S(x)$ and ${\cal S}$ can be expressed in terms of
incomplete elliptic integral of the second kind $E(qx|-(qA)^2)$ as:
\begin{eqnarray}
\int_0^x \sqrt{1+\psi_u^2}du &=& \int_0^x du \sqrt{1+(qA)^2\textrm{sen}^2 qu}\nonumber\\ &=&
\frac{1}{q}E(qx|-(qA)^2)
\end{eqnarray}
Therefore, the roughness parameters $r$ and $r_c$ can be expressed as:
\begin{eqnarray}
r&=&\frac{2E(qL/2|-(qA)^2)}{qL}=\frac{2}{\pi}E(-(qA)^2)\\
r_c&=&\frac{E(qx_c|-(qA)^2)}{qx_c}
\end{eqnarray}
where $E(x)$ is the complete elliptic integral of the second kind, and where
$x_c$ can be obtained from (\ref{filling-equil-cond}) as:
\begin{equation}
x_c=\frac{\pi-\arcsin\left(\frac{\tan \theta}{qA}\right)}{q}
\label{xc}
\end{equation}
This solution only exists if $\tan \theta<qA$. Under these conditions,
it is easy to see that there is another solution to (\ref{filling-equil-cond}):
\begin{equation}
x_c^*=\frac{\arcsin\left(\frac{\tan \theta}{qA}\right)}{q}
\label{xc2}
\end{equation}
which corresponds to a maximum of free energy ${\cal F}^{F}$. This can be seen from the behavior
free energy ${\cal F}^{D}$ taking $x_c\equiv x$ as a free parameter in the range $[0,L/2]$.
For small $x$, we can see from (\ref{free-energy-filled-2}) that ${\cal F}^{F}(x)\approx {\cal F}^{D}+
{\cal A} \sigma_ {lv} (1 - \cos \theta) qx/\pi$, and therefore it is an increasing function at $x=0$. 
On the other hand, for $x\approx L/2$,
${\cal F}^{F}(x) \approx {\cal F}^{W}+{\cal A} \sigma_ {lv}(1-\cos \theta) (qx/\pi-1)$, which is also an increasing function
at $x=L/2$. Therefore, since $x_c^*<x_c$ and by continuity of the free energy function ${\cal F}^{F}(x_c)$,
we conclude that $x_c$ given by (\ref{xc}) must correspond to a minimum of ${\cal F}^{F}(x_c)$, and $x_c^*$ given
by (\ref{xc2}) to a maximum. Figure \ref{fig2} shows graphically the behavior of
${\cal F}^{F}(x)$ for different situations. As mentioned above,
the $F$ state exists only if $\theta<\theta^*=\arctan(qA)$, which corresponds to the spinodal of this state.
By decreasing $\theta$ (increasing temperature), the free energy of the filled state, i.e. ${\cal F}^F(x_c)$, decreases
until reaches the value of ${\cal F}^{D}$ for $\theta=\theta_f$ at the filling transition. As a consequence, the 
macroscopic theory predicts that the filling transition must be first-order. On the other hand, 
note that ${\cal F}^{W}>{\cal F}^{F}$ for $\theta>0$. This observation has two consequences. First, the filling
transition occurs at lower temperatures than the complete wetting temperature predicted by the Wenzel law.  
As $\theta$ is further decreased, the thermodynamic equilibrium state corresponds to $F$. So,
filling transition preempts Wenzel complete wetting transition.
The value of $x_c$ given by (\ref{xc}) increases as $\theta$ decreases and reaches the value of $x_c=L/2$ for 
$\theta=0$. Therefore, the macroscopic theory predicts that the wetting transition on a sinusoidal substrate
is \emph{continuous} and occurs at the same temperature that for the the flat substrate. Therefore the existence of 
first-order wetting transitions (and associated off-coexistence transitions such as prewetting) cannot be predicted
by the macroscopic theory. 

\begin{figure}[t]
\centerline{\includegraphics[width=0.7\columnwidth]{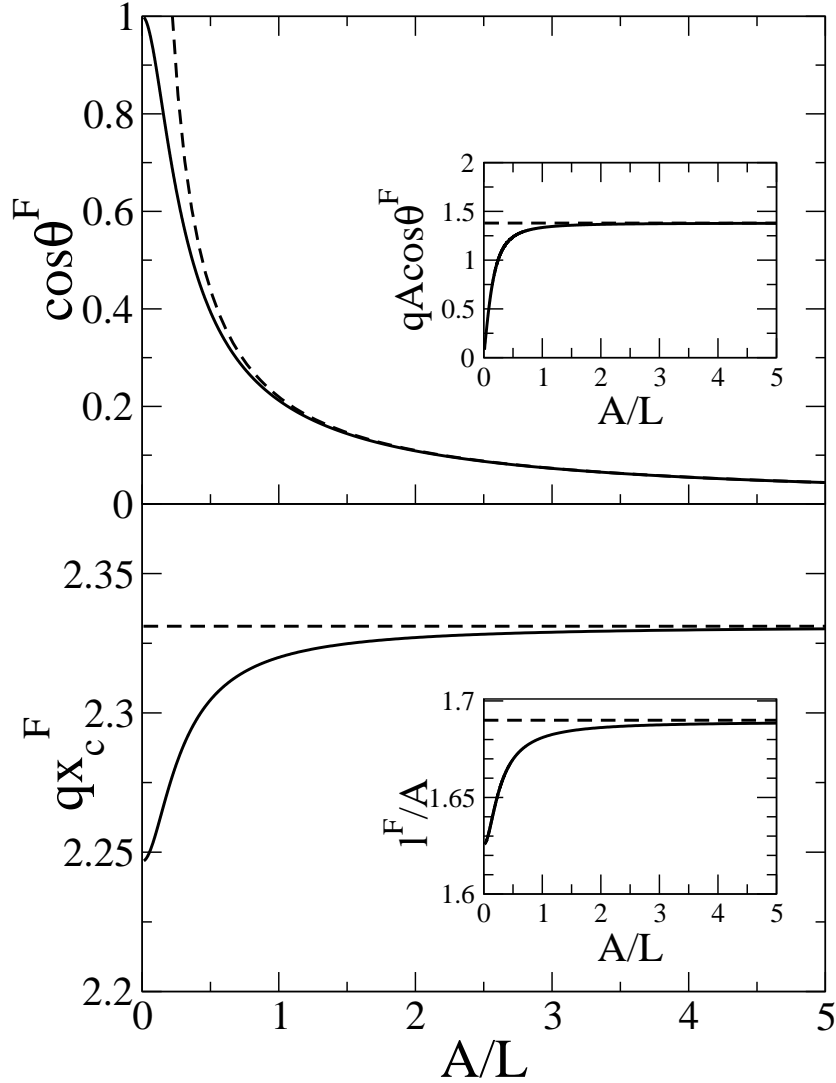}}
\caption
{Top panel: plot of the cosine of the contact angle at the filling transition $\theta^F$ for $h=0$ 
as a function of $A/L$. The dashed line corresponds to the asymptotic expression (\ref{thetafrough}). Inset, representation
of $qA\cos\theta^F$ as a function of $A/L$, being the dashed line the limiting value from the asympotic analysis for
rough substrates. Bottom panel: plot of $qx_c^F$ as a function of $A/L$. The dashed line is the limiting value
from the asympotic analysis. Inset: plot of the midpoint interfacial height above the substrate of the $F$ state at
the filling transition, $l^F$, in units of the substrate amplitude $A$, 
as a function of $A/L$. The dashed line corresponds to the asymptotic value for $A/L\to \infty$.}
\label{fig3}
\end{figure}

Figure \ref{fig3} shows the dependence of the contact angle at filling transition, $\theta^F$, and the value of 
$x_c$ at the filling transition, $x_c^F$, as a function of $A/L$. We see that, for large $A/L$, $\cos \theta^F$ scales
as $L/A$, while $qx_c^F$ is quite insensitive to the value of $A/L$ and asymptotically tends to a constant as
$A/L\to \infty$. To explain this behaviour, recall that the values of $\theta^F$ and $x_c^F$ solve simultaneously
(\ref{filling-equil-cond}) and (\ref{df21}). For large $(qA)^2$, we can approximate $\sqrt{1+\psi_x^2}\approx |\psi_x|=
qA\sin qx$ for $x>0$. Thus, (\ref{filling-equil-cond}) leads to the condition:
\begin{equation}
\sin qx_c^F=\frac{1}{qA\cos\theta^F}
\label{fecrough}
\end{equation}
which is compatible with (\ref{xc}) if $\theta\approx \pi/2$. Substituting this expression in (\ref{df21}), we reach 
to the following equation for $x_c^F$:
\begin{equation}
qx_c^F \sin qx_c^F + \cos qx_c^F = 1
\label{xcfrough}
\end{equation}
with a solution $qx_c^F\approx 2.33$. Consequently, the midpoint interfacial height for rough substrates
is almost independent of $L$, and approximately equal to $A(1-\cos qx_c^F)\approx 1.69A$.
Substituting (\ref{xcfrough}) into (\ref{fecrough}), we have the following asymptotic
expression for $\theta^F$:
\begin{equation}
\cos\theta^F \approx 0.22 \frac{L}{A}
\label{thetafrough}
\end{equation}
We see from figure \ref{fig3} that these asymptotic expressions are extremely accurate for values of $A/L>1$.

To finish the description of the macroscopic theory, we note that both $D$ and $F$ states can also be obtained out of
the two-phase coexistence. In the case of the $F$ states, they can observed on a limited range of values of chemical
potencial close to coexistence, where the liquid is still a metastable state. 
Their typical configurations are shown in figure \ref{fig1}(iv): the liquid-vapour 
interface is no longer flat but shows a cylindrical shape, with a radius given by the Young-Laplace equation:
\begin{equation}
R=\frac{\sigma_{lv}}{(\rho_l-\rho_g)|\Delta \mu|}
\label{laplace}
\end{equation}
where $\rho_l$ and $\rho_g$ are the liquid and vapour densities at coexistence, and $\Delta \mu$ is the chemical potential
shift with respect to the coexistence value. The free energy ${\cal F}^F$ is obtained by making a suitable modification
of (\ref{free-energy-filled-2}) as:
\begin{eqnarray}
{\cal F}^{F}&=&{\cal F}^{D}+\frac{2{\cal A}}{L}\sigma_{lv}\Bigg(\frac{R}{2}\arcsin\frac{x_c}{R}\nonumber\\
&-&\cos\theta\int_0^{x_c}\sqrt{1+\psi_x^ 2(x)}dx 
+\frac{x_c\psi(x_c)}{R}\nonumber\\
&-&\frac{1}{R}\int_0^{x_c} dx \psi(x)+\frac{x_c}{2}\sqrt{1-\left(\frac{x_c}{R}\right)^2}\Bigg)
\label{free-energy-filled-3}
\end{eqnarray}
At the equilibrium configuration the liquid-vapour interface makes contact 
with the substrate at a value $x=x_c$ where the angle between the liquid-vapour interface and the substrate is equal to
the contact angle for the flat substrate $\theta$, so $x_c$ is the solution of the following implicit equation: 
\begin{equation}
\theta=\arctan\psi_x (x_c)-\arcsin\frac{x_c}{R}
\label{outcoexistence1}
\end{equation}
For the sinusoidal substrate, this equation reads:
\begin{equation}
\theta=\arctan (qA\sin qx_c)-\arcsin\frac{x_c}{R}
\label{outcoexistence2}
\end{equation} 
which can be solved numerically or graphically, with a solution $qx_c$ which is a function of $\theta$, $qA$ and $qR$. 

The off-coexistence filling transition occurs when ${\cal F}^{F}={\cal F}^{D}$. By using (\ref{free-energy-filled-3}) and 
(\ref{outcoexistence1}), it is possible to find numerically the characteristics of the $F$ state which is at equilibrium 
with the $D$ state. For the sinusoidal substrate, we find that $qx_c$ is a function only of $qA$ and $qR$. Our numerics show
that, for fixed $qA$, the midpoint interfacial height $l$, defined as:
\begin{equation}
l=A(1-\cos qx_c)-R(1-\sqrt{1-(x_c/R)^2})
\label{outcoexistence3}
\end{equation}
decreases as $qR$ decreases (i.e. $|\Delta \mu|$ increases), until vanishes for some critical value of $qR$. This state 
corresponds to the macroscopic theory prediction for the critical point of the filling transition.   

\section{Methodology\label{sec3}}

\begin{figure}[t]
\centerline{\includegraphics[width=\columnwidth]{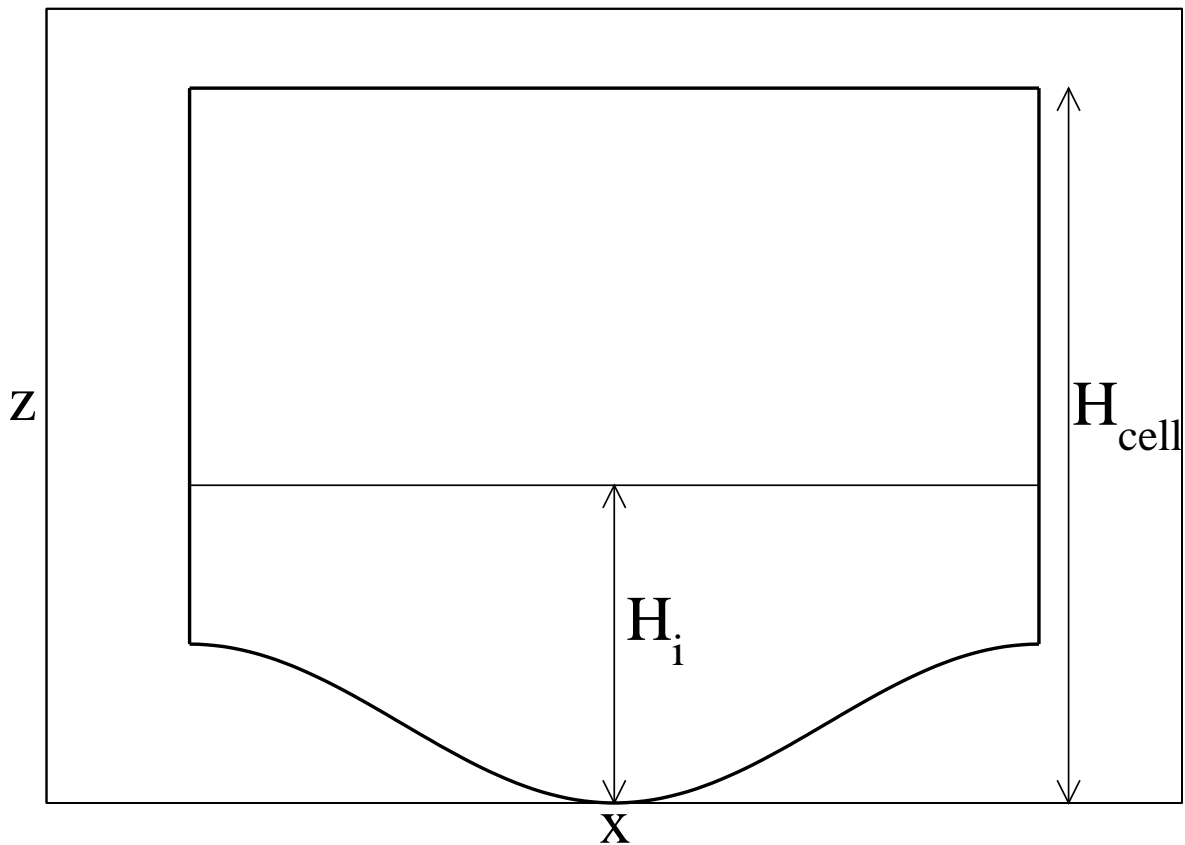}}
\caption
{Finite-size geometry considered in our numerical study of the sinusoidal substrate. See text for explanation.}
\label{fig4}
\end{figure}
Our starting point is the Landau-Ginzburg functional for subcritical temperatures:
\begin{eqnarray}
{\cal F}&=&\int_{V} d\mathbf{r} \left[\frac{1}{2}({\bnabla}m)^{2}- h m
+\frac{1}{8}(m^2 - 1)^2\right]\nonumber\\&+&\int_{S} d\mathbf{s}
\frac{c}{2}(m-m_{s})^2
\label{redfree-energy3}
\end{eqnarray}
based on a magnetization order-parameter $m(\mathbf{r})$. As explained in the Appendix, with this choice the bulk 
magnetization at coexistence takes the value $1$ or $-1$, and the bulk correlation length $\xi=1$, which provides the 
unit of length for all length scales. Taking into account the continuous translational symmetry along the $y$-axis,
and the periodicity across the $x$-axis, the minimization of the functional (\ref{redfree-energy3}) is performed in
the geometry depicted in figure \ref{fig4}. The bottom boundary is one period of the sinusoidal substrate shape 
(\ref{defpsi}). The values of the magnetization at this boundary are free, except in the case $c=\infty$, where
the Dirichlet boundary condition $m(\mathbf{s})=m_{s}$ is imposed. 
On the top boundary at $z=H_{cell}$, the magnetization is fixed to the bulk value ($-1$ if $h=0$). 
The value of $H_{cell}$ must be large enough in order to mimic the effect of an infinite domain. We take as value
of $H_{cell}=4A$, for which we did not find any size-effect. Finally, periodic boundary conditions are imposed at the 
vertical boundaries.   

We have numerically minimized the free-energy functional (\ref{redfree-energy3}) to determine the equilibrium magnetization
profiles for different substrate geometries and surface couplings. The minimization was done with a finite element
method, using a conjugate-gradient algorithm to perform the minimization. The numerical discretization of the continuum 
problem was performed with adaptive triangulation coupled with the finite-element method in order to resolve different
length scales \cite{patricio0}. This method was succesfully applied to the minimization of a Landau-de Gennes functional 
for the study of interfacial phenomena of nematic liquid crystals in presence of microstructured substrates 
\cite{patricio1,patricio2,patricio3,patricio4,patricio5}. For each substrate geometry and value of surface enhancement $c$,
we obtain the different branches of interfacial states $D$, $F$ and $W$ on a wide range of values of the surface
coupling (either $m_{s}$ or $h_1$) for the bulk ordering field $h=0$. 
Additionally, in order to locate the off-coexistence filling transition and, when the wetting transition is first-order, 
the prewetting line, we also explored the different free-energy branches out of coexistence, i. e. $h<0$. In this case,
the values of the surface coupling are restricted to be above the filling and wetting transitions at bulk coexistence, i.e. 
$h=0$. The true equilibrium state will be the state that gives the least free energy at the same thermodynamic conditions, 
and the crossing between the different free-energy branches will correspond to the phase transitions. 
Finally, the interface will be localized by using a crossing criterion, i. e. at the points where the order parameter 
profile vanishes. 

The initial state for each branch is obtained at a suitable value of the 
surface coupling by using as initial condition for the minimization procedure a state where the magnetization profile takes
a constant value $+1$ for the mesh nodes with $z<H_i$ (see figure \ref{fig4}), and $-1$ otherwise. After minimization, 
the mesh is adapted and the functional is minimized again. We iterate this procedure a few times (typically 2-4 times). 
The value of $H_i$ depends on the branch: $H_i=0$ for the $D$ branch (i.e. the initial magnetization profile is $-1$ 
everywhere), $H_i\sim A+H_{cell}/2$ for the $W$ branch and $H_i\gtrsim A$ for the $F$ branch. Once the first state
is obtained, we may follow the branch slowly modifying the value of the surface coupling, using as initial condition
for the next value of the surface coupling the outcome corresponding to the current minimization.   
Alternatively to the procedure outlined above, we may obtain the $W$ free-energy branch for $h=0$ 
by imposing a fixed value of the
magnetization $+1$ on the top boundary, and using as initial magnetization profile $+1$ everywhere. In order to obtain
the free energy of the $W$ states, we add to the minimized free energies the contribution due to an interface between the
two bulk coexisting phases, which is equal to $2L/3$ (see (\ref{sigma+-})), where $L$ is the period of the sinusoidal
substrate in the $x$-axis.

\section{Numerical results\label{sec4}}

Following the methodology described in the previous section, we numerically studied the interfacial phenomenology
that the system shows in presence of the sinusoidal substrate within the mean-field approximation.
As interfacial Hamiltonian theories point out that the phenomenology will 
depend on the type of wetting transition when the system is in contact with a flat substrate \cite{Rascon,Rejmer2}, 
we consider two situations: $c=0$ and $c=+\infty$, that correspond to first-order and critical wetting, respectively
(see appendix). Theory predicts the ratio between the amplitude and roughness period $A/L$ is a key parameter, so 
in general we consider the cases $A/L = 0.5,1,1.5$ and $2$, although for some systems we have considered other values of 
$A/L$. To assess the finite-size effects, for each value of $A/L$ we consider different substrate periods in a range 
$L=5-100$.  

\subsection{Results for $c=0$, $h=0$}

Under this condition, the relevant surface coupling parameter is $h_1$, taken as the limit $c\to 0$, $ṃ_{s}\to \infty$
and $c m_{s}\to h_1$. Furthermore, the surface coupling energy in (\ref{redfree-energy3}), up to an irrelevant 
constant, has the expression $-\int_{\cal S} d\mathbf{s} h_1 m(\mathbf{s})$. As shown in the Appendix, the
reduced surface coupling $h_1$ plays the role of the temperature $T$, as $h_1\sim (T_c-T)^{-1}$, where $T_c$ is the bulk 
critical temperature. The first-order wetting transition for a flat substrate occurs for a surface field $h_1=h_1^{w,\pi}
\approx 0.34$. On the other hand, the prewetting critical point occurs at $(h_1,h)=(h_1^{cpw},h^{cpw})
\approx (0.847,-0.1925)$. Therefore, we have explored the values of $h_1\in [0,1]$ and $h\in [-0.5,0]$. 

We start our study under bulk coexistence conditions, i.e. $h=0$. In order to compare the minimization results for $h=0$
with the macroscopic theory, we obtained analytical expressions for the
free-energy densities $f\equiv {\cal F}/{\cal A}$, where ${\cal F}$ is the interfacial free energy and ${\cal A}$ is the 
projected area
of the substrate in the $x-y$ plane. Substitution of the Landau-Ginzburg surface tensions (\ref{sigmaw-3}), (\ref{sigma+-})
and (\ref{sigmaw+}) into (\ref{free-energy-dry}), (\ref{free-energy-filled}) and (\ref{free-energy-wet}) leads to the 
following expressions for the three free-energy branches at $h=0$:
\begin{eqnarray}
f^{D}=\frac{2}{3\pi}\left(1-(1-2h_1)^{3/2}\right)E(-(qA)^2)
\label{fi}\\
f^{F}=
\frac{2}{3}-\frac{2}{3\pi}
\beta \nonumber\\
+\frac{2}{3\pi}(1-(1-2h_1)^{3/2})E(-(qA)^2)
\label{fii}\\
-\frac{1}{3\pi}\Big((1+2h_1)^{3/2}-(1-2h_1)^{3/2}\Big)E\left(\pi-\beta |-(qA)^2\right) 
\nonumber\\
\mbox{with } \beta=\arcsin\frac{1}{qA}\sqrt{\frac{4}
{\left((1+2h_1)^{3/2}-(1-2h_1)^{3/2}\right)^2}-1}
\nonumber\\
f^{W}=\frac{2}{3}+\frac{2}{3\pi}\left(1-(1+2h_1)^{3/2}\right)
E(-(qA)^2)
\label{fiii}
\end{eqnarray}
where $ E(x)$ and $ E(x|y)$ are the complete and incomplete elliptic integrals
of the second kind, respectively, and $qA=2\pi A / L $. 
\begin{figure}[t]
\centerline{\includegraphics[width=\columnwidth]{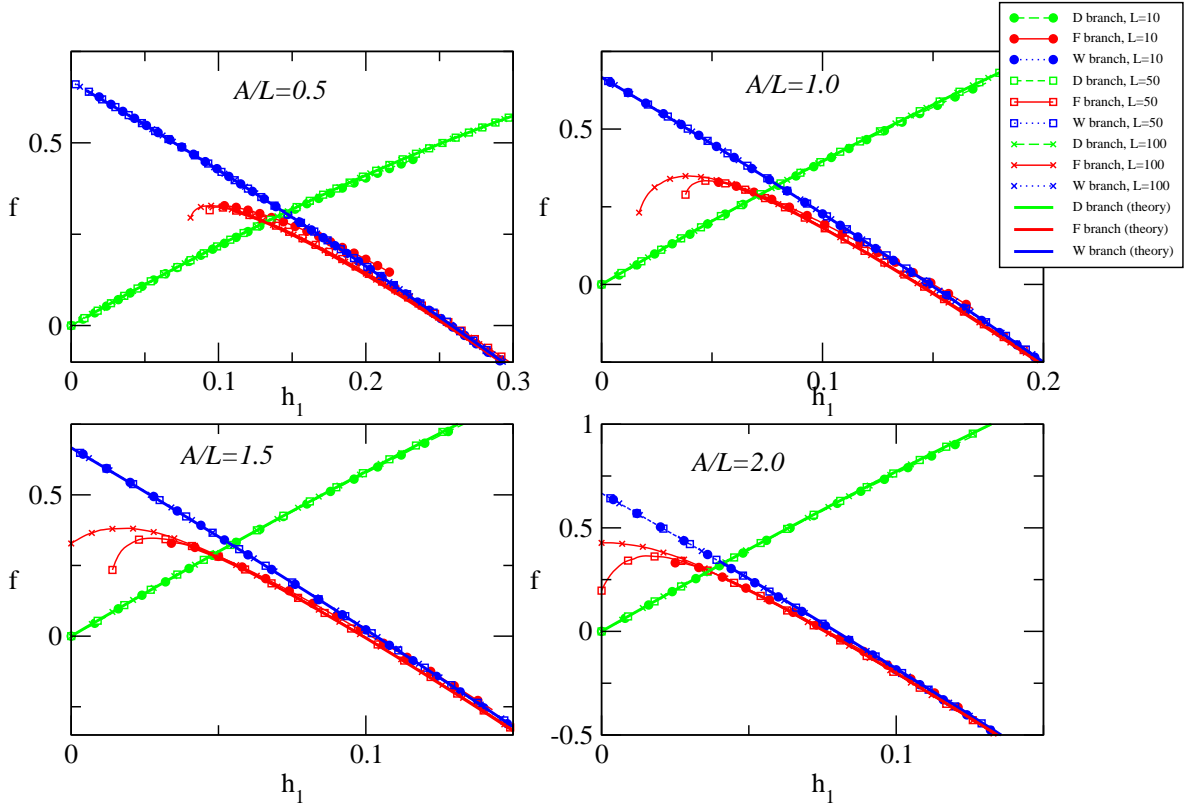}}
\caption
{Plot of the free energy densities of the different branches of interfacial states for $c=0$ at $h=0$, 
as a function of the surface field $h_1$, for a sinusoidal substrate with $A/L=0.5, 1, 1.5$ and $2$, and $L=10$ 
(crosses), $L=50$ (open squares) and $L=100$ (filled circles). The $D$ states branch corresponds to the green (lighter grey)
symbols, the $W$ states branch to the blue (dark grey) symbols and the $F$ states branch to the red (light grey) symbols. 
For comparison, the theoretical prediction from (\ref{fi}), (\ref{fii}) and (\ref{fiii}) are also represented as continuous 
lines (the colour code is the same as for the numerical results).}
\label{fig5}
\end{figure}

\begin{figure}[tbp]
\centerline{\includegraphics[width=\columnwidth]{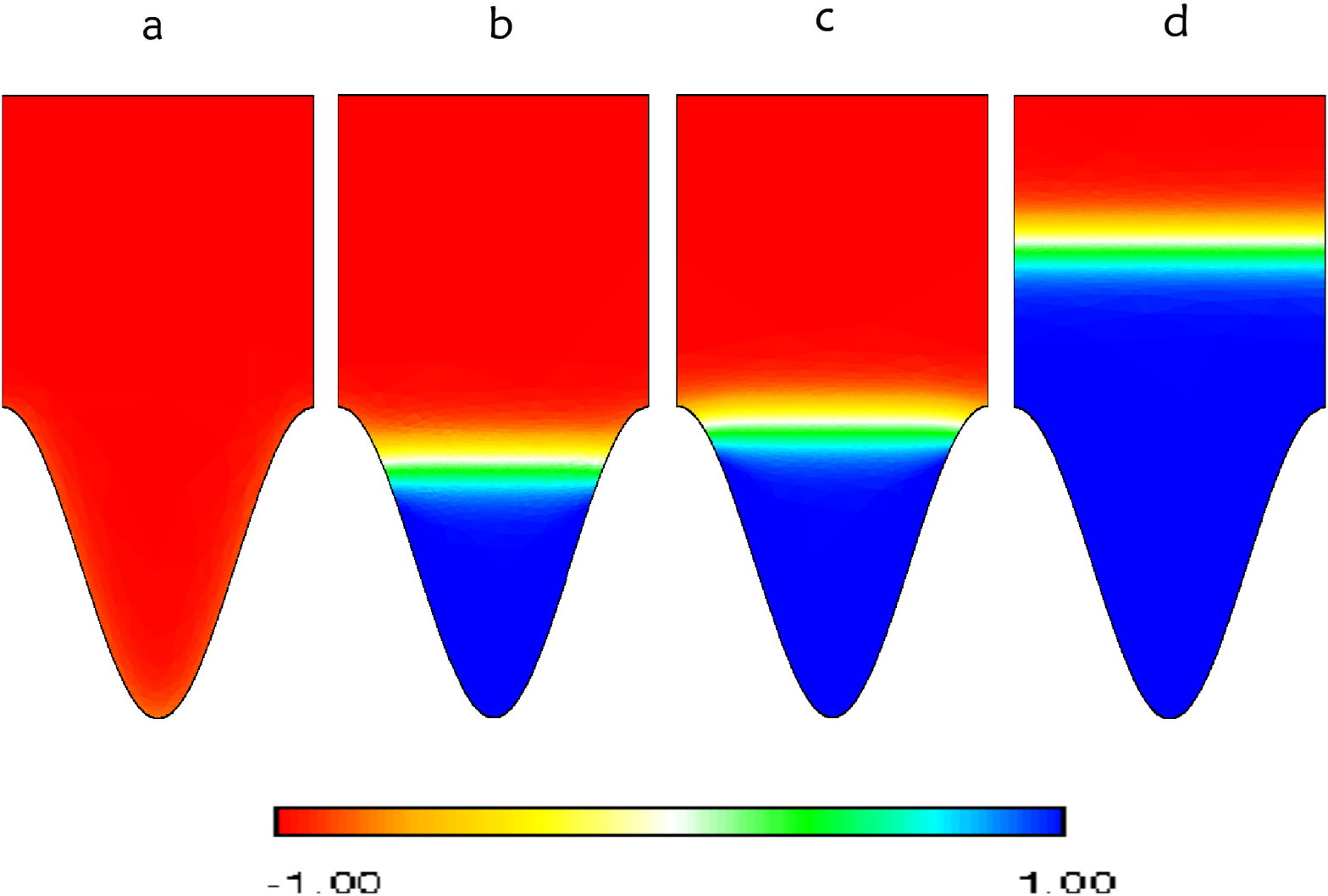}}
\caption{Magnetization profiles corresponding to the 
coexistence states at the filling transition (a and b) and the wetting transition (c and d)
for $c=0$, $h=0$, $A/L=0.5$ and $L=20$.}
\label{fig6}
\end{figure}

Figure \ref{fig5} shows the free energy densities of the different branches as a function of $h_1$ at bulk coexistence.
For a fixed value of $A/L$, the $D$ and $W$ branches are quite insensitive to the substrate periodicity $L$, and 
converge quickly to the macroscopic expressions (\ref{fi}) and (\ref{fiii}). On the other hand, the $F$ branch is more
sensitive to $L$, specially for the largest values of $h_1$, although also converges to the macroscopic expression 
(\ref{fii}) for moderate values of $L$. 
For small values of $A$ the free energy density of the $F$ branch exceeds the limiting value given by (\ref{fii}), and if 
$A$ is of order of the correlation length the $F$ branch becomes metastable in all the range of values of $h_1$ with 
respect to $D$ or $W$ states. In this situation, there is only a 
first-order wetting transition between a $D$ and a $W$ state located at the value of $h_1$ given by the Wenzel law. 
But in general, filling and the wetting transitions are located as the intersection between the $D$ and $F$ branches, and
the $F$ and $W$ branches, respectively. Thus these transitions are both first-order. However, although the filling 
transition is clearly first-order in all the cases, the first-order character of the wetting transition weakens as $L$ is 
increased. Figure \ref{fig6} shows the typical magnetization profiles at the filling and wetting transition.   
We can see that the coexisting magnetization profiles at the filling transition are in good agreement with the schematic
picture shown in figure \ref{fig1}, and the mid-point interfacial height follows accurately the macroscopic prediction.
On the other hand, at the wetting transition (which for the macroscopic theory is continuous), we see that the mid-point
interfacial height at the $F$ state (c) is slightly below the substrate maximum height $2A$. This fact may indicate that
the wetting transition of the rough substrate for large $L$ is controlled by the wetting properties of the substrate 
at its top, with corrections associated to the substrate curvature there. If this hypothesis is correct, then
the wetting transition should remain first-order for all $L$ and converge asymptotically to the wetting transition of 
the flat wall as $L\to \infty$. We will come back to this issue below.

\begin{figure}[tbp]
\centerline{\includegraphics [width = \columnwidth]{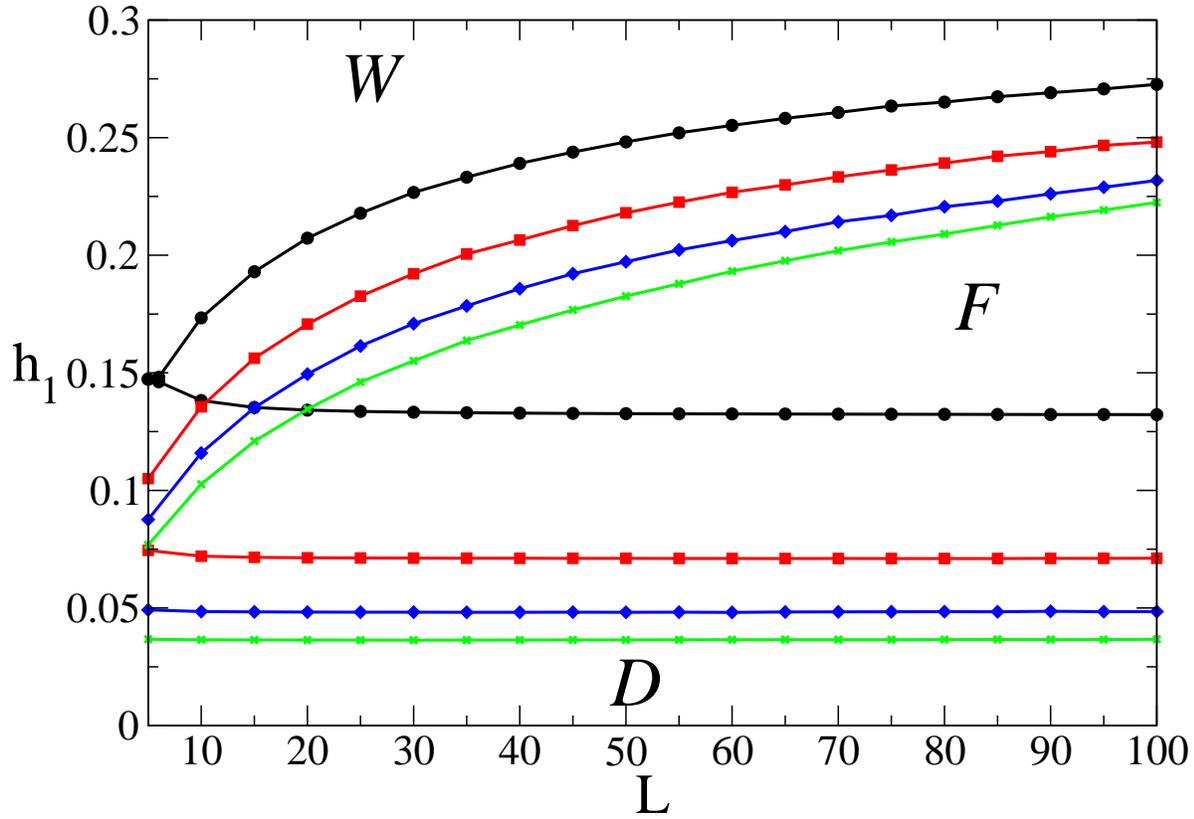}}
\caption
{Adsorption phase diagram on a sinusoidal substrate with $c=0$ and $h=0$. The phase boundaries
between $D$, $F$ and $W$ states are plotted for $A/L=0.5$ (black circles), $A/L=1$ (red squares), $A/L=1.5$ (blue diamonds)
and $A/L=2$ (green crosses). The lines serve only as guides for the eyes. 
}
\label{fig7}
\end{figure}

Figure \ref{fig7} represents the adsorption phase diagram at bulk coexistence. The phase boundaries correspond
either to filling transitions between $D$ and $F$ phases, or wetting transitions between either a $F$ or $D$ phase and 
a $W$ phase. We can see that the substrate roughness enhances the wettability of the substrate: as the substrate is
rougher the wetting and filling transitions are shifted to lower values of $h_1$, leading to an increase of the
stability region of the $W$ phase at the expense of the $F$ phase, and a reduction of the stability region of the $D$ 
phase with respect to the $F$ phase. 

For the shallowest considered substrate $A/L=0.5$ we see that for very small values of $L$ there is only a first-order 
wetting transition between a $D$ and a $W$ state at a value of $h_1$ almost independent of the value of $L$ given by
Wenzel law prediction. As $L$ increases the $F$ phase appears at a triple point for $L\approx 6$ from which the filling
and wetting transition (between $F$ and $W$ states) emerge. We have checked that, as the substrate becomes shallower,
this triple point occurs for larger values of $L$: $L\approx 25$ for $A/L=0.2$ and $L>100$ for $A/L=0.1$. In all the cases,
the value of $A\sim 5-10$. On the contrary, for larger values of $A/L$ we do not observe this scenario
in the range of values of $L$ studied, but we expect to observe it for smaller values of $L$. In any case, 
for moderate and large values of $A$, we see that the filling transition line is almost independent of the value of $L$ 
and it coincides with the macroscopic theory prediction. In particular, from (\ref{thetafrough}) and as 
$\cos\theta\approx 3h_1$ for small $\theta$, the filling transition value is approximately equal to 
$0.073\times(L/A)$ for $A/L\ge 1$. On the other hand, the wetting 
transition has a strong $L$-dependence, so the transition value of $h_1$ increases with $L$. It is worthwhile to note
that these wetting transition values are always smaller than the corresponding one to the flat substrate 
$h_1^{w,\pi}\approx 0.34$, in agreement with the predictions from interfacial Hamiltonian theory \cite{Rejmer,Kubalski}. 

\begin{figure}[tbp]
\centerline{\includegraphics [width = \columnwidth]{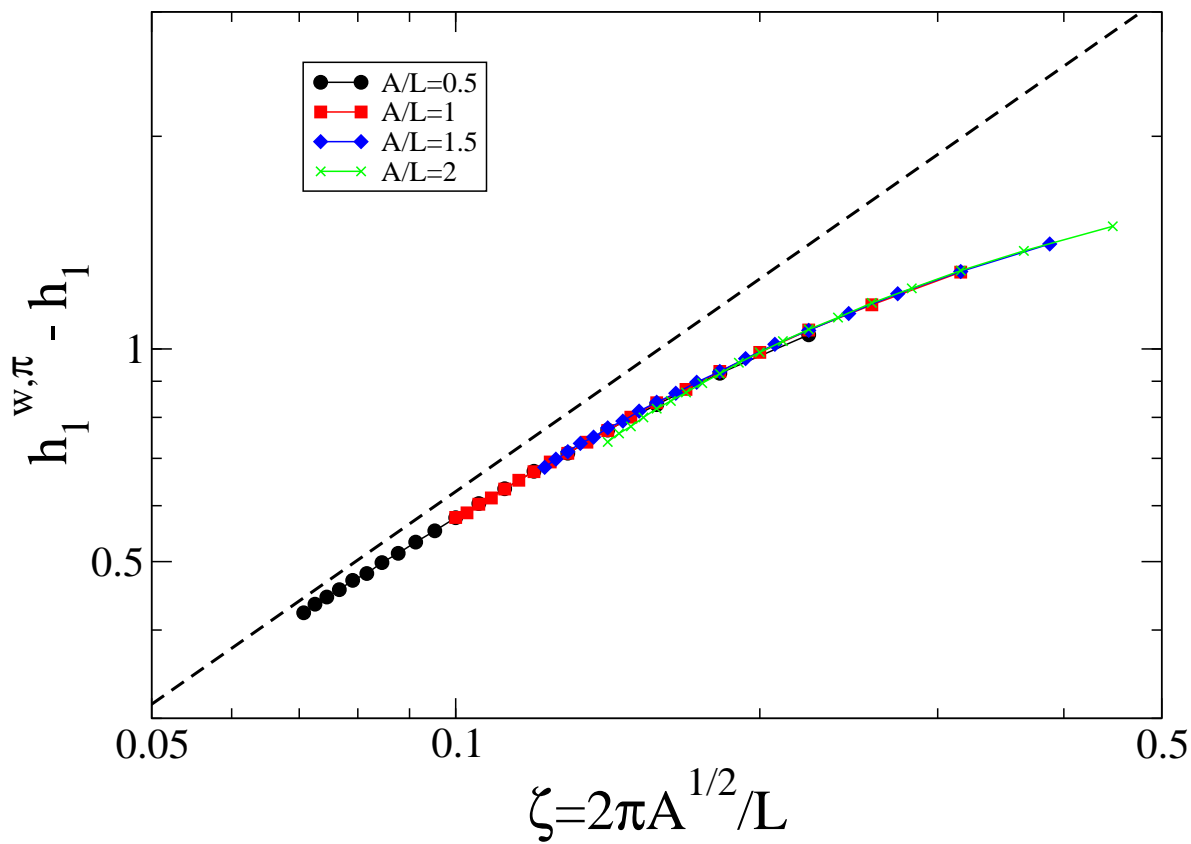}}
\caption
{Plot of the wetting transition shift $h_1^{w,\pi}-h_1$ with respect to the flat substrate as a function of the
curvature of the substrate at its top $\zeta=2\pi\sqrt{A}/L$ for $L=10-100$ and $A/L=0.5$ (circles), $A/L=1$ 
(squares), $A/L=1.5$ (diamonds) and $A/L=2$ (crosses). The dashed line indicates a linear dependence of the
wetting transition shift with $\zeta$. 
}
\label{fig8}
\end{figure}

In order to check the hypothesis mentioned above that the wetting transition for large $L$ is just a curvature-driven 
correction to the wetting transition of the flat substrate at the top of the substrate, we plot in figure \ref{fig8} the
wetting transition shift with respect to the flat value $h_1^{w,\pi}-h_1$ as a function of $\zeta = q \sqrt{A}$, which
is the square root of the curvature at the substrate top. 
Our numerical data show a fairly good collapse in a master curve. For small $\zeta$, this master curve seems to show an 
asympotically linear dependence with $\zeta$. A simple argument may rationalize this result. Recall that close to the
wetting transition the $F$ state is characterized by an almost flat gas-liquid interface at a height slightly
below the maximum substrate height $2A$. Consequently, the free-energy difference between the $F$ and $W$ states $\Delta
{\cal F}$ comes
from contribution of the region close to the substrate maximum. If $\zeta$ is small, we may approximate the shape of 
the substrate by the parabolic approximation $\psi(x) \sim 2A - \zeta^2 (x-L/2)^2/2$. We can assume that the interfacial
height with respect to the substrate maximum is close to the corresponding for the flat substrate for the partial wetting 
phase at the wetting transition. So, there will be a contribution to $\Delta {\cal F}$ which is proportional to the 
free-energy difference between the partial and complete wetting interfacial states at the wetting transition, which 
is proportional to $h_1-h_1^{w,\pi}$ close to the transition, and to the length of the segment in the $x-$axis 
where there is no interface in the $F$ state, which is inversely proportional to $\zeta$. Obviously this contribution 
is always negative if $h_1<h_1^{w,\pi}$. Thus, there must be another contribution to $\Delta {\cal F}$ 
which takes into account the distorsions in the magnetization profile with
respect to the flat situation driven by the substrate curvature. This contribution should be positive, and we can assume 
that it is nearly constant for small $\zeta$. At the wetting transition for the rough substrate, $\Delta {\cal F}$ should
vanish. So, from the balance between these two terms of $\Delta {\cal F}$, we conclude that at the wetting transition 
$h_1-h_1^{w,\pi} \sim \zeta$. Our observations seem to support this argument, but results for smaller values of $A$ 
and/or larger values of $L$ should be needed in order to establish its validity beyond any doubt. 
  
\subsection{Results for $c=+\infty$, $h=0$}
\begin{figure}[t]
\centerline{\includegraphics[width=\columnwidth]{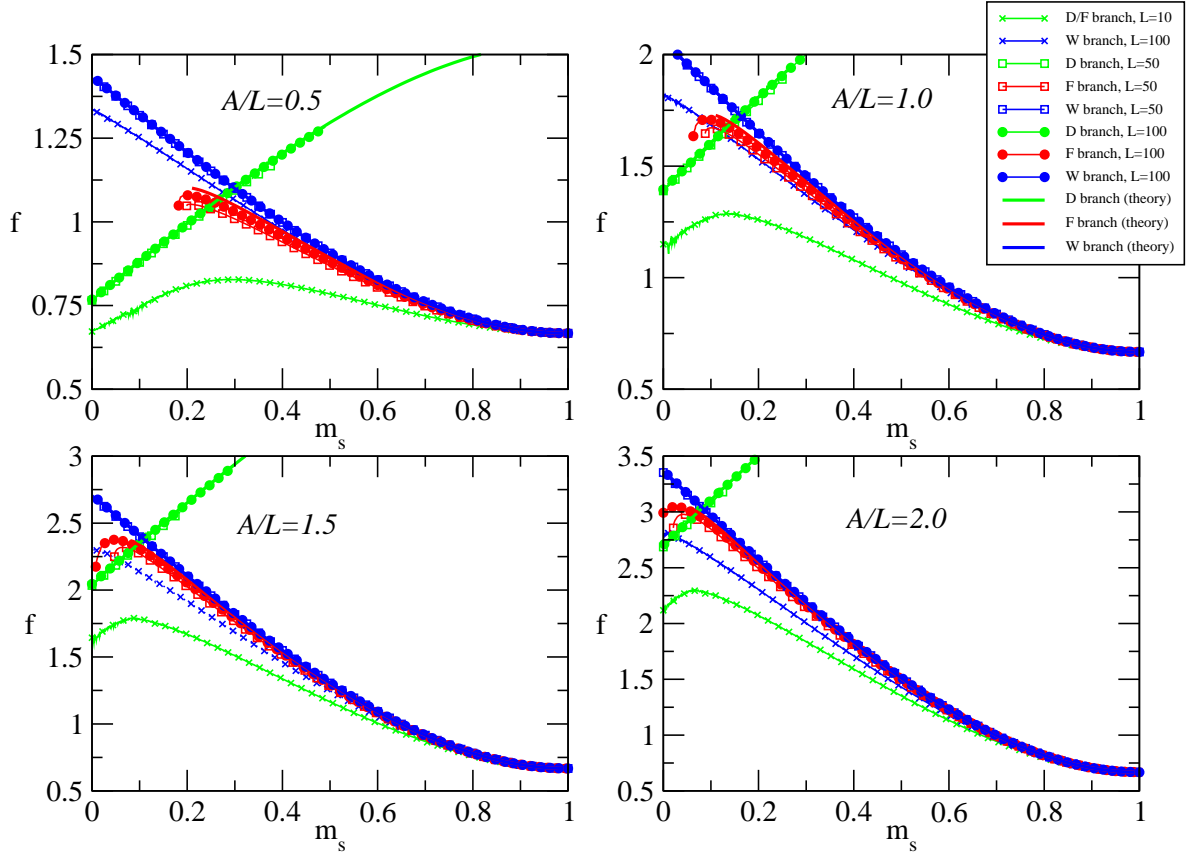}}
\caption
{Plot of the free energy densities of the different branches of interfacial states for $c=\infty$ at $h=0$, 
as a function of the surface magnetization $m_{s}$, for a sinusoidal substrate with $A/L=0.5, 1, 1.5$ and $2$, and $L=10$ 
(crosses), $L=50$ (open squares) and $L=100$ (filled circles). The $D$ states branch corresponds to the green (lighter grey)
symbols, the $W$ states branch to the blue (dark grey) symbols and the $F$ states branch to the red (light grey) symbols. 
For comparison, the theoretical prediction from (\ref{fi-2}), (\ref{fii-2}) and (\ref{fiii-2}) are also represented as 
continuous lines (the colour code is the same as for the numerical results).}
\label{fig9}
\end{figure}
\begin{figure}[tbp]
\centerline{\includegraphics [width = \columnwidth]{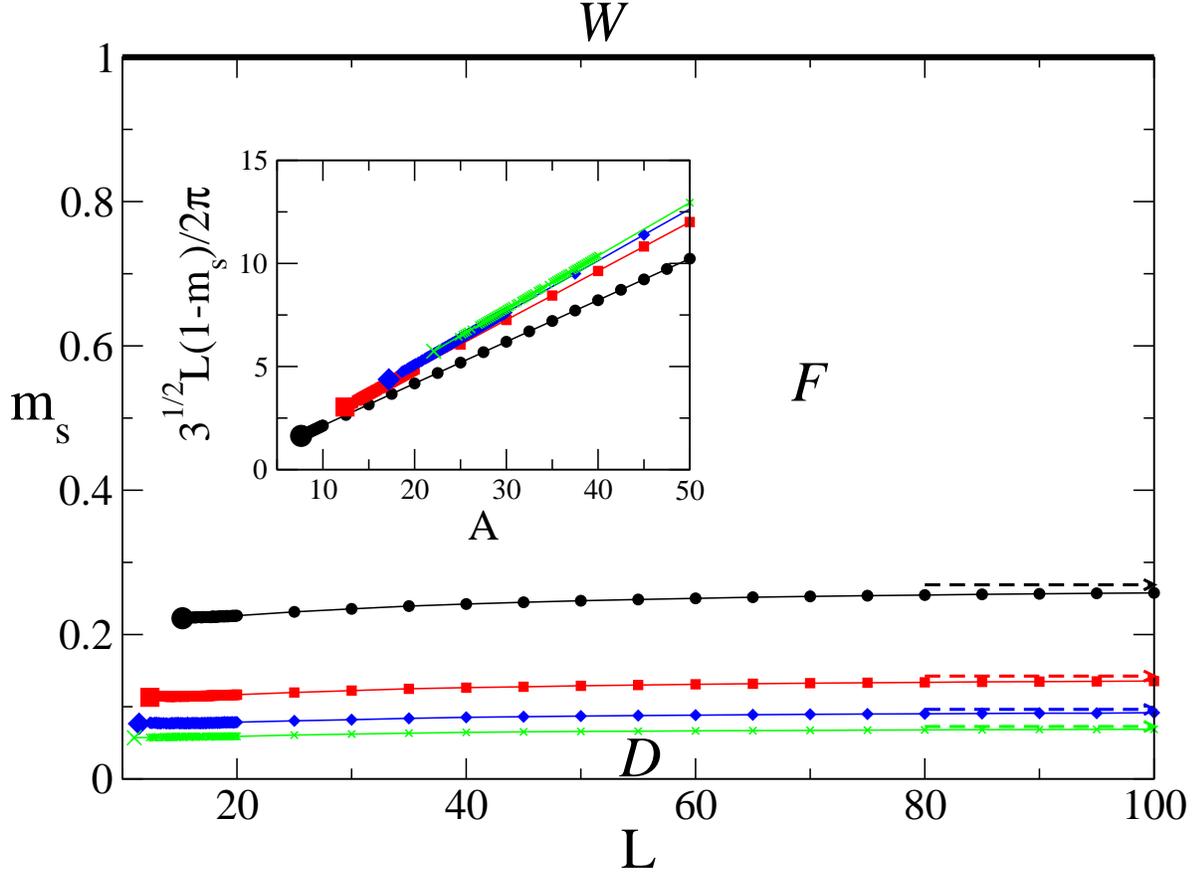}}
\caption
{Adsorption phase diagram on a sinusoidal substrate with $c=\infty$ and $h=0$. The phase boundaries
between $D$ and $F$ states are plotted for $A/L=0.5$ (black circles), $A/L=1$ (red squares), $A/L=1.5$ (blue diamonds)
and $A/L=2$ (green crosses) (the lines serve only as guides for the eyes). The arrows correspond to the filling transition
values of $m_{s}$ predicted by the macroscopic theory, and the big symbols to the filling transition critical points.
Finally the wetting transition is represented by the thick continuous line for $m_s=1$. Inset: plot of $\Delta \tilde{T}=
\sqrt{3}L(1-m_s)/(2\pi)$ as a function of the substrate amplitude $A$ along the filling transition line. 
The meaning of the symbols is the same as in the main plot.
}
\label{fig10}
\end{figure}

When the \emph{enhancement} parameter tends to infinity, we can drop the surface coupling energy in
(\ref{redfree-energy3}), but the magnetization at the surface is fixed to the value $m_{s}$. 
As shown in the Appendix, the surface magnetization $m_{s}$ plays the role of the temperature $T$, as $m_{s}\sim 1/\sqrt{T_c-T}$, where $T_c$ is the bulk critical temperature. The system in contact with a flat substrate has a critical 
wetting transition when the surface order parameter $m_{s}\to 1$. Therefore, we proceed in a similar way to the case 
$c=0$, so the reduced free energy (\ref{redfree-energy3}) is minimized subject to Dirichlet boundary conditions at the
substrate for values of $m_{s}$ between 0 and 1 and the bulk ordering field $h\in [-0.5,0]$.

We start with the bulk coexistence conditions, i.e. $h=0$. Figure \ref{fig9} represents 
the free energy densities as a function of $m_{s}$ for the branches $D$, $F$ and $W$.
As in the case $c=0$, each figure corresponds to a fixed value of $A/L$ and we consider
different values of $L$ to assess the finite-size effects. We also plot the theoretical 
predictions obtained from the macroscopic approach, which would correspond to the $L\to \infty$ 
limit:
\begin{eqnarray}
f^{D}&=&\frac{2}{\pi}\left(\frac{m_{s}}{2}-\frac{m_{s}^3}{6}+
\frac{1}{3}\right)E(-(qA)^2)
\label{fi-2}\\
f^{F}&=&
\frac{2}{3}-\frac{2}{3\pi}
\beta\nonumber\\
&+&\frac{2}{\pi}\left(\frac{m_{s}}{2}-\frac{m_{s}^3}{6}+
\frac{1}{3}\right)E(-(qA)^2)
\label{fii-2}\\
&-&\frac{1}{3\pi}\left(3m_{s}-m_{s}^3\right)
E\left(\pi-\beta
|-(qA)^2\right)
\nonumber\\
\mbox{with}& &\beta=\arcsin\left(\frac{1}{qA}\sqrt{\frac{4}{\left(3m_{s}-m_{s}^3\right)^2}-1}\right)
\nonumber\\
f^{W}&=&\frac{2}{3}+\frac{2}{\pi}\left(-\frac{m_{s}}{2}+\frac{m_{s}^3}{6}
+\frac{1}{3}\right) E(-(qA)^2)
\label{fiii-2}
\end{eqnarray}
These results show several differences compared to the case $c=0$. 
First, for every $A/L$ finite-size effects on $L$ are more pronounced in all branches, specially 
in the $F$ branch. On the other hand, for small values of $L$ the filling transition disappears
as there is a continuous crossover from $D$ to $F$ states. 
Finally, the $W$ branch is always metastable in the range $m_{s}\in [0,1]$, and touches 
tangentially the $F$ branch at $m_{s}=1$.  
In fact, we observe a continuous unbinding of the interface along the $F$ branch as $m_{s}\to 1$ 
from the magnetization profiles. For $m_{s}>1$, the $F$ and $W$ branches coincide. 
From these observations we conclude that the wetting transition at the rough
substrate is always continuous, and at the same value $m_{s}=1$ as in the flat substrate. 
This is in agreement with the predictions of interfacial Hamiltonian theories \cite{Rascon,Rejmer2}.
On the other hand, the filling transition shows a more pronounced finite-size dependence on $L$ than in the case $c=0$.
Figure \ref{fig10} shows the adsorption phase diagram at $h=0$ for different values of $A/L$. 
For a fixed value of $A/L$ and large $L$, the filling transition value of $m_{s}$ increases with $L$, although it is
bounded from above by the macroscopic theory transition value. Furthermore, the filling transition shifts towards 
lower values of $m_{s}$ as the substrate is rougher. As in the $c=0$ case, by using (\ref{thetafrough}) 
and taking into account that $\cos\theta\approx 3m_{s}/2$ for small $m_{s}$, we find that that the limiting
value for $m_{s}$ at the filling transition scales as $0.146\times (L/A)$ for $A/L\ge 1$.
As $L$ decreases, the filling transition disappears at a critical point, so for smaller values of $L$ we observe
the continuous crossover between the $D$ and $F$ states. The existence of this critical point was also 
observed in the framework of interfacial Hamiltonian theories \cite{Rascon,Rejmer2}. Furthermore, these theories
also predict that along the filling transition line, the rescaled temperature $\Delta \tilde{T}$ is a function of
$A$, regardless the value of $L$ \cite{Rascon}. By using (\ref{bindingpotential}), in our case 
$\Delta \tilde{T}=\sqrt{3}L(1-m_s)/(2\pi)$. The inset of the
figure \ref{fig10} shows that, although the filling transition lines for different values of $A/L$ seem to converge
for small $A$, they deviate as the substrate amplitude increases. This observation is consistent with
the fact that the interfacial Hamiltonian theories are valid in the shallow substrate limit. 

\begin{figure}[tbp]
\centerline{\includegraphics[width=\columnwidth]{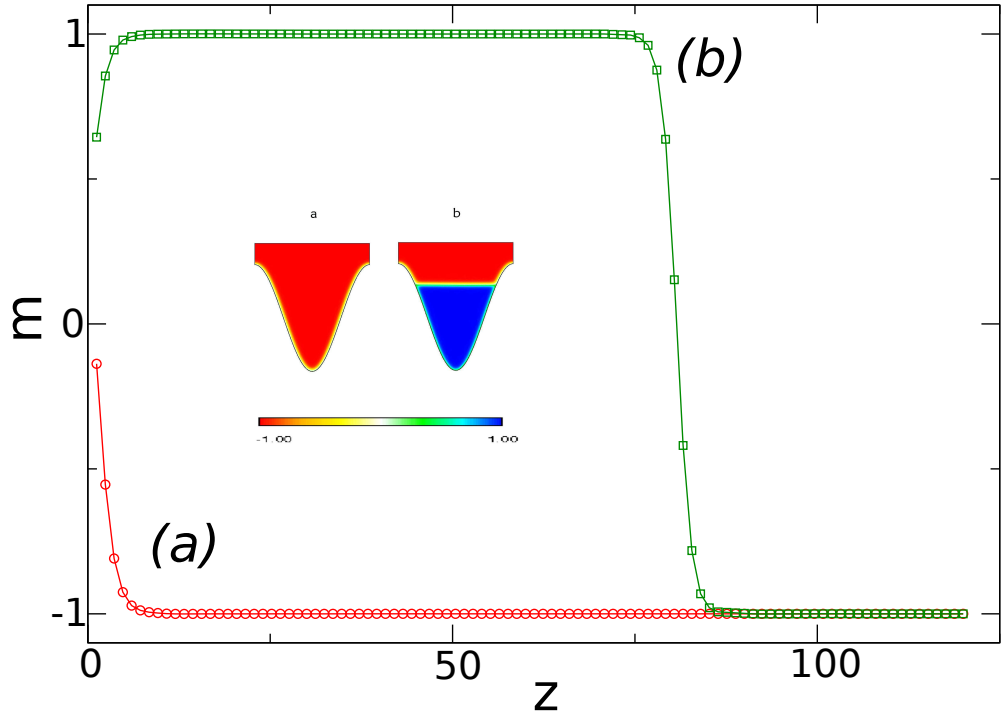}}
\caption
{Magnetization profile along the vertical axis at $x=0$ corresponding to: (a) the $D$ state and (b) the $F$ state
at coexistence in the filling transition for $c=\infty$, $A/L=0.5$ and $L=100$. Inset: plot of the
complete magnetization order parameter profiles of the coexisting $D$ and $F$ states.}
\label{fig11}
\end{figure}

\begin{figure}[tbp]
\centerline{\includegraphics[width=\columnwidth]{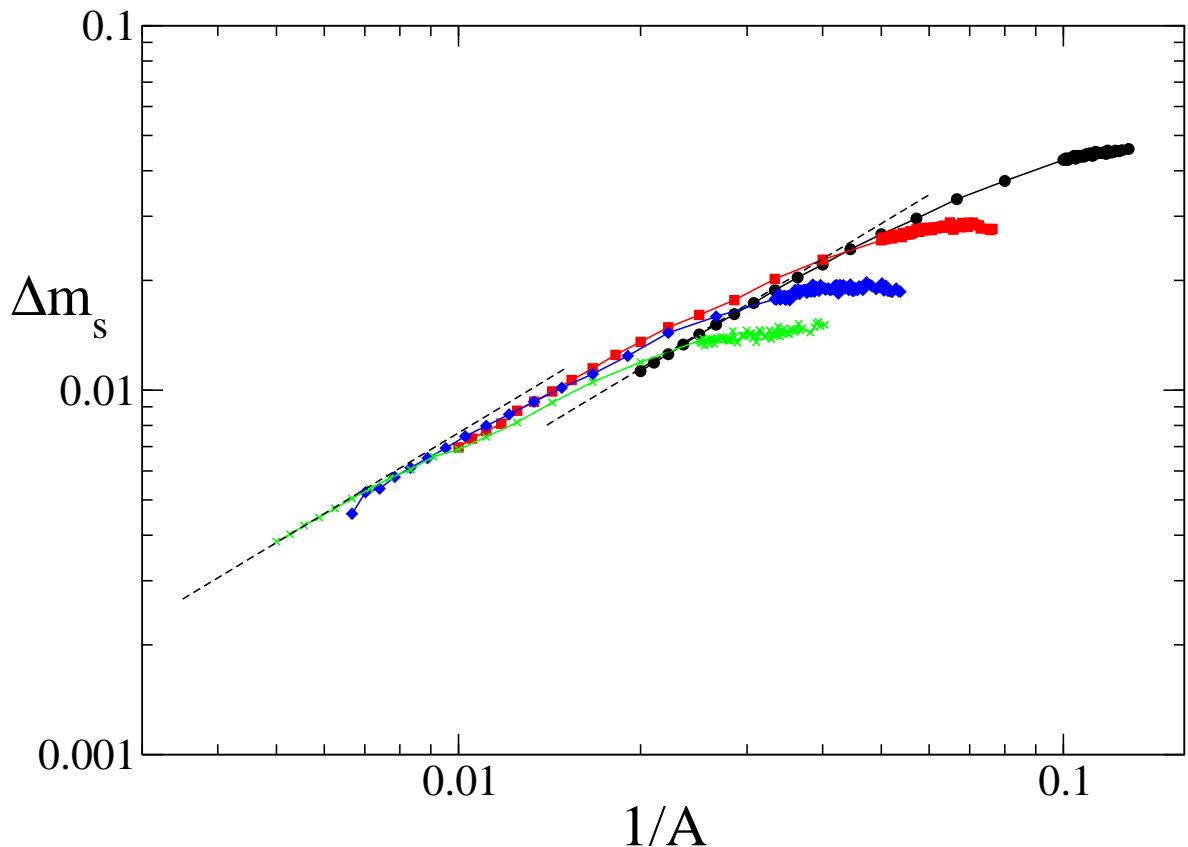}}
\caption
{Plot of the filling transition shift $m_{s}^{macro}-m_{s}$ as a function of $1/A$, for $A/L=0.5$ (circles),
$A/L=1$ (squares), $A/L=1.5$ (diamonds) and $A/L=2.0$ (crosses). Dashed lines correspond to the linear fits for
the transition values for large $A$ and $A/L=0.5$ and $A/L=2$.
}
\label{fig12}
\end{figure}

In order to characterize the filling transition, we choose as the order parameter of the interface position
along the vertical $x=0$, i.e. above the minimum of the substrate. Figure \ref{fig11} 
plots two typical magnetization profiles at the filling transition. 
The position of the interface is determined as the height at which the magnetization profile vanishes.
If the magnetization profile is always negative (as in the $D$ state in figure \ref{fig11}), the interfacial height
is undetermined. These results show that the profiles are in agreement with the picture outlined in figure
\ref{fig1}. So, any finite-size dependence of the transition value of $m_{s}$ for large $L$ with respect to the
macroscopic prediction must arise from the order parameter profile distortions induced by the regions where the 
interface touches the substrate. The correction to $f^F$ associated to these distortions scales as $B/L$, where
$B$ is the line tension associated to the liquid-vapor-substrate triple line and which depends on the contact angle $\theta$.
So, we expect that the shift of the transition value $m_{s}$ with respect to the macroscopic prediction 
$m_{s}^{macro}$ should scale as $1/A$ for large $L$ and fixed substrate roughness $A/L$. 
Our results shown in figure \ref{fig12} are in agreement with this prediction. Furthermore, we observe that the shift
becomes almost independent of $L$ for the roughest substrates $A/L \ge 1$. In order to explain this result, we may
expand the free-energy density around $m_{s}^{macro}$ for large $A/L$,  and keeping the leading order terms, 
we obtain that, at filling transition:
\begin{equation}
f_F - f_D  \approx -3.38\frac{A}{L}(m_{s}-m_{s}^{macro}) + \frac{B}{L}=0 
\label{df3}
\end{equation}  
from which $m_{s}^{macro}-m_{s}\propto 1/A$.

\begin{figure}[tbp]
\centerline{\includegraphics[width=\columnwidth]{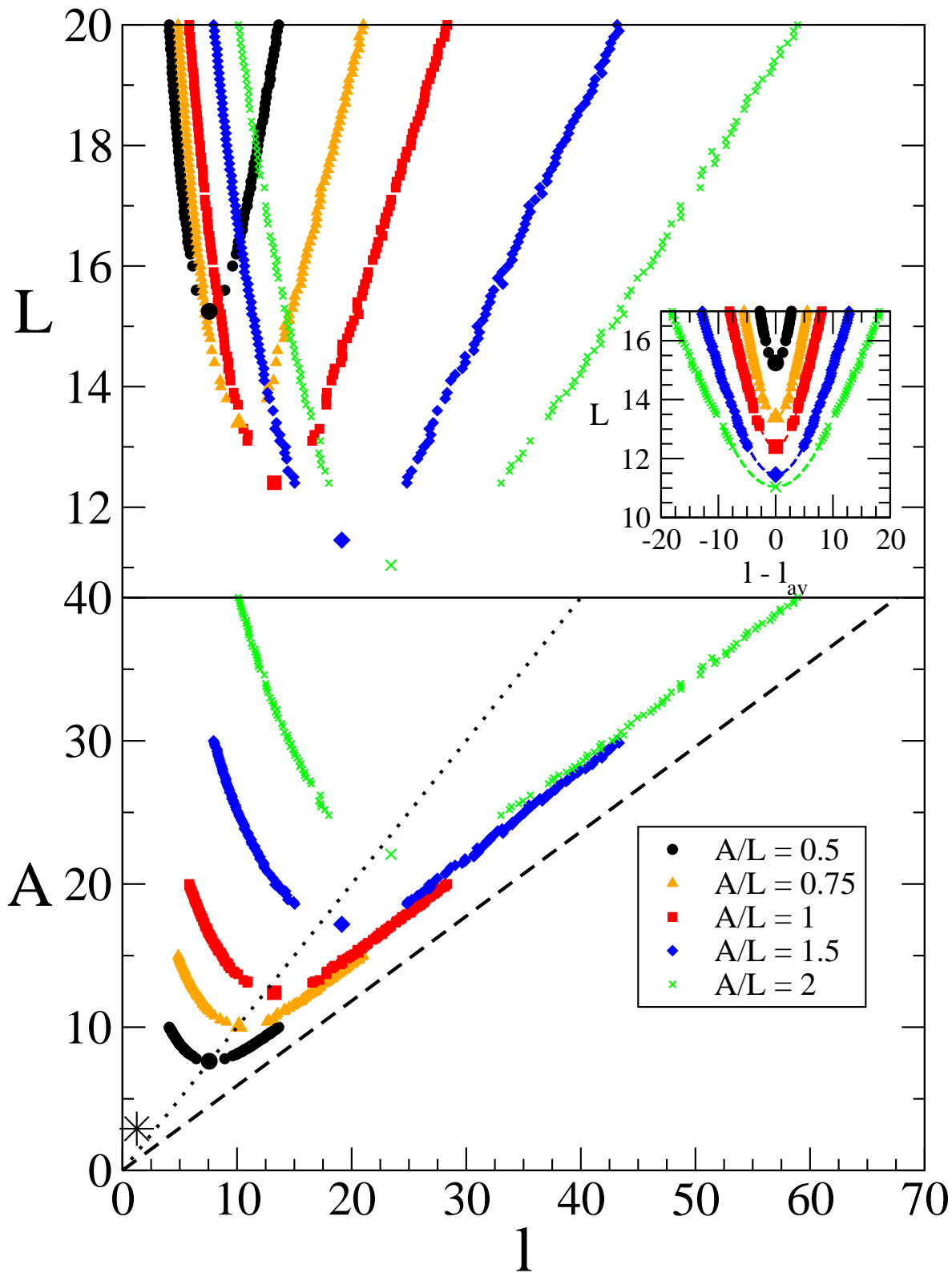}}
\caption
{Plot of the mid-point interfacial height $l$ of the coexisting $D$ and $F$ states at the filling transition 
for $h=0$ as a function of $L$ (top) and $A$ (bottom) close to its critical point, for $A/L=0.5$ (circles), 
$A/L=0.75$ (triangles), $A/L=1$ (squares), $A/L=1.5$ (diamonds) and 
$A/L=2.0$ (crosses). The star in the bottom panel corresponds to the location of the filling critical point 
predicted for shallow substrates from the interfacial Hamiltonian approach \cite{Rascon}, and the dashed and dotted lines
in the bottom panel are the mid-point interfacial height of the $F$ state predicted from the macroscopic
theory for large $A/L$ at the filling transition and the spinodal line of the $F$ states, respectively. 
Inset: plot of the deviations of the mid-point interfacial height of the $D$ and $F$ states at 
the filling transition for $h=0$ with respect to their average value (the meaning of the symbols is the same as in the
main panel).}
\label{fig13}
\end{figure}

For small $L$, the filling transition ends up at a critical point. Figure \ref{fig13} plots the behaviour of 
the mid-point interfacial height of the coexisting $D$ and $F$ states, $l_D$ and $l_F$, respectively, 
at the filling transition under bulk coexistence conditions. As a function of $A$, the mid-point interfacial height 
of the $F$ states show a weak dependence on $L$, but its value is below the macroscopic prediction for large $A/L$, 
$l\approx 1.69 A$. Close to the filling transition critical point our numerical scheme is not very accurate, so we are
not able to locate directly the critical point. In order to estimate the location of the filling transition critical
point, we followed a procedure very similar to the used to locate usual bulk liquid-gas transitions. First, we evaluate
the average value $l_{av}=(l_F+l_D)/2$ of the interfacial heights of the coexisting $D$ and $F$ states for each $h_1$. 
After that, we substract to the interfacial heights $l_D$ and $l_F$ the average value computed previously (see inset
of figure \ref{fig13}). This curve is quite symmetric around zero. Finally, we fit to a parabola the values of $l_D-l_{av}$
and $l_F-l_{av}$ for small values of $L$ (i.e. close to the critical point), so the parabola height at its maximum gives an
estimate of the critical value of $L$, and the value of $l_{av}$ at the critical $L$ gives the corresponding midpoint
interfacial height. From figures \ref{fig10} and \ref{fig13} we see that, for the rougher substrates the critical
value of $L$ slightly decreases as $A/L$ increases. On the other hand, by decreasing $A/L$ the increase of the critical 
value of $L$ is steeper. Regarding the critical values of $A$, we see that they increase as the substrate roughness 
increases, being this dependence nearly linear for the roughest substrates, in agreement with the fact that the critical
value of $L$ depends weakly on $A/L$ for rough substrates. The location of the critical filling points is close
to the spinodal line of the $F$ states obtained from the macroscopic theory ($qx_c = \pi/2$ and $l=A$).  
Finally, it is worth to note that, if we extrapolate to the
shallow substrate limit, i.e. $A/L\to 0$, our results are compatible with the predictions of interfacial Hamiltonian 
theories for shallow substrates, where the critical amplitude is $A=2.914$, independently of the value of $L$ \cite{Rascon}.
However, our results show that this prediction is no longer valid for rougher substrates.

\subsection{Results for $c=0$, $h<0$}

\begin{figure}[tbp]
\centerline{\includegraphics[width=\columnwidth]{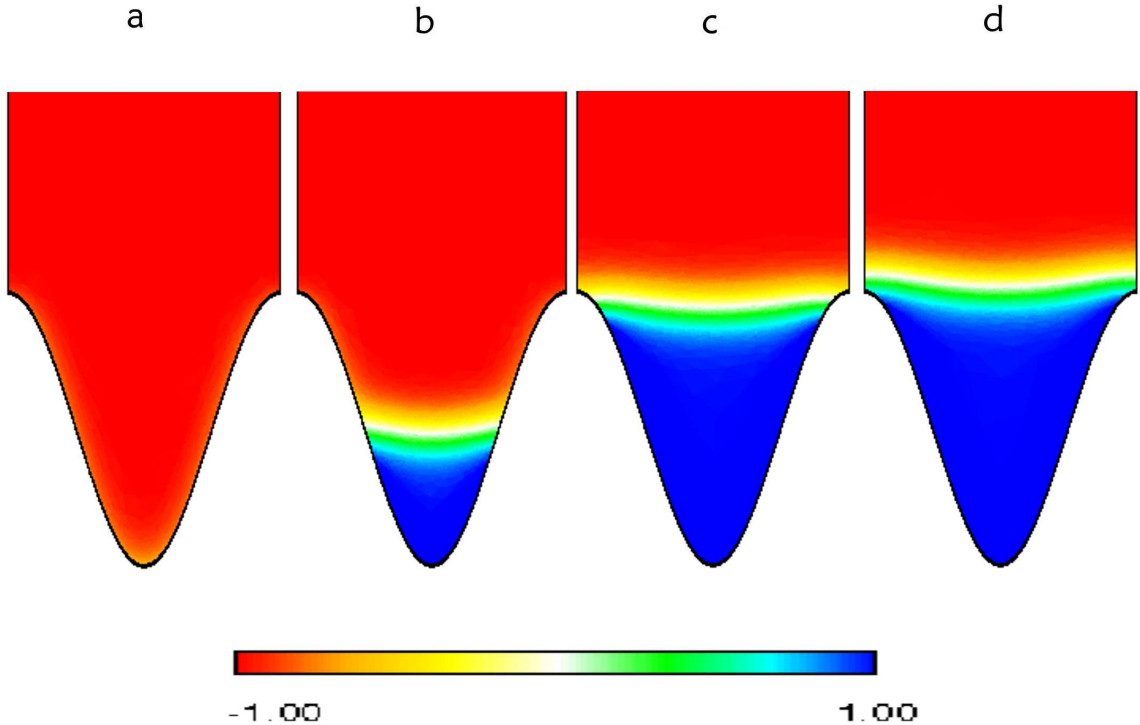}}
\caption{Magnetization profiles corresponding to the 
coexistence states at the filling transition (a and b) for $h=-0.025$ and the wetting transition (c and d) for
$h=-0.0105$ for $c=0$, $A/L=0.5$ and $L=20$.}
\label{fig14}
\end{figure}

We turn back to the case $c=0$, and now we explore the bulk off-coexistence interfacial phenomenology. In order to keep
the bulk phase with negative magnetization as the true equilibrium state, we consider that the ordering field $h<0$.  
Typically, we observe three different interfacial states with a finite adsorption, which are the continuation inside 
the off-coexistence region of the $D$, $F$ and $W$ branches, and that we will denote as $D$, $F$ and $F^*$ states, 
respectively. The $D$ states are very similar to their counterparts at $h=0$, except close to the filling critical
point (see below), where a small adsorbed region of liquid develops on the substrate groove. The $F$ states show partially
filled grooves, where the liquid-vapour interface is curved, as shown in figure \ref{fig1}(iv). Finally, the $F^*$ states
correspond to completely filled grooves, with a thicker microscopic layer of liquid on top of the substate, which
diverges as $h\to 0$. As in the $F$ state, typically the liquid-vapour interface in the $F^*$ states is curved. 

\begin{figure}[t]
\centerline{\includegraphics[width=\columnwidth]{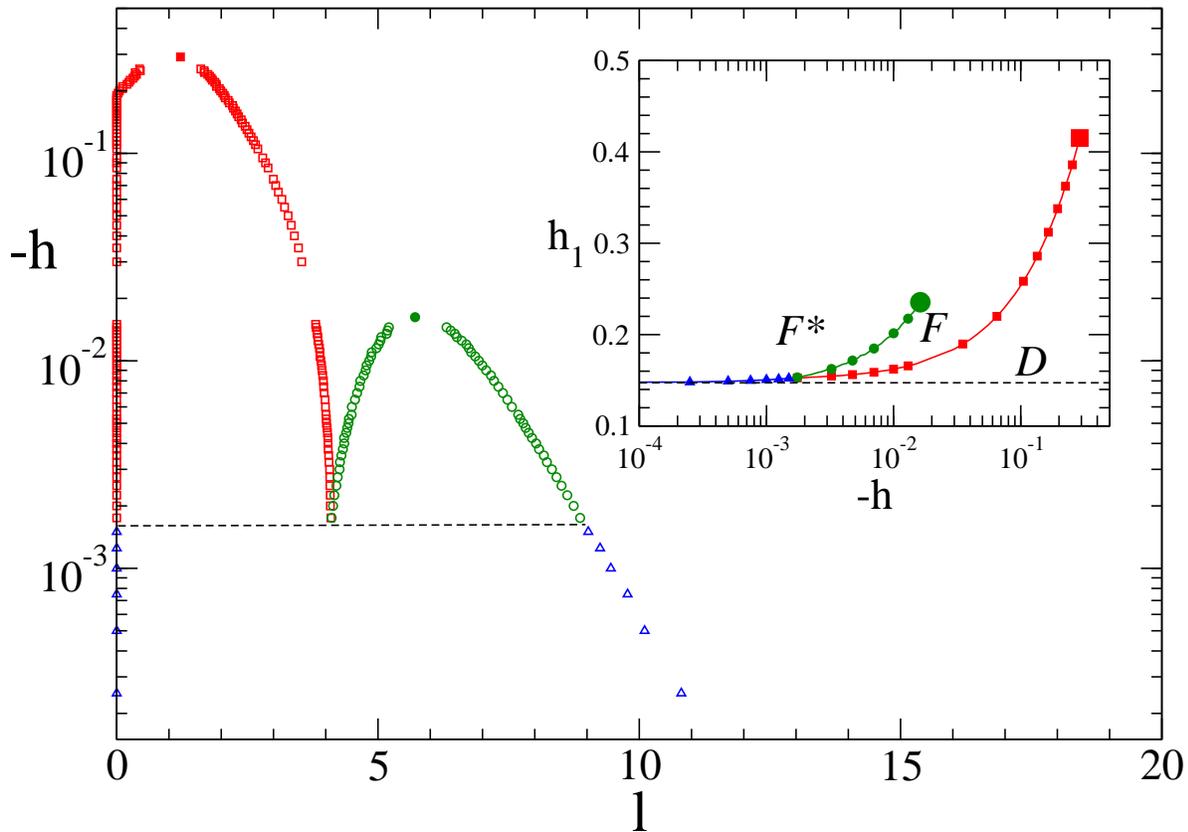}}
\caption{Plot of the midpoint interfacial height of the coexisting $D$ and $F$ states at filling transition 
(red squares), $F$ and $F^*$ states at the prewetting transition (green circles) and $D$ and $F^*$ states at 
the Wenzel prewetting transition (blue triangles). The full symbols correspond to the filling and prewetting 
critical points, and the dashed line indicates the location of the $D-F-F^*$ triple point. Inset: off-coexistence 
interfacial phase diagram. The meaning of the symbols is the same as in the main panel. The dashed line indicates the
location of the wetting transition at bulk coexistence.}
\label{fig15}
\end{figure}

\begin{figure}[t]
\centerline{\includegraphics[width=\columnwidth]{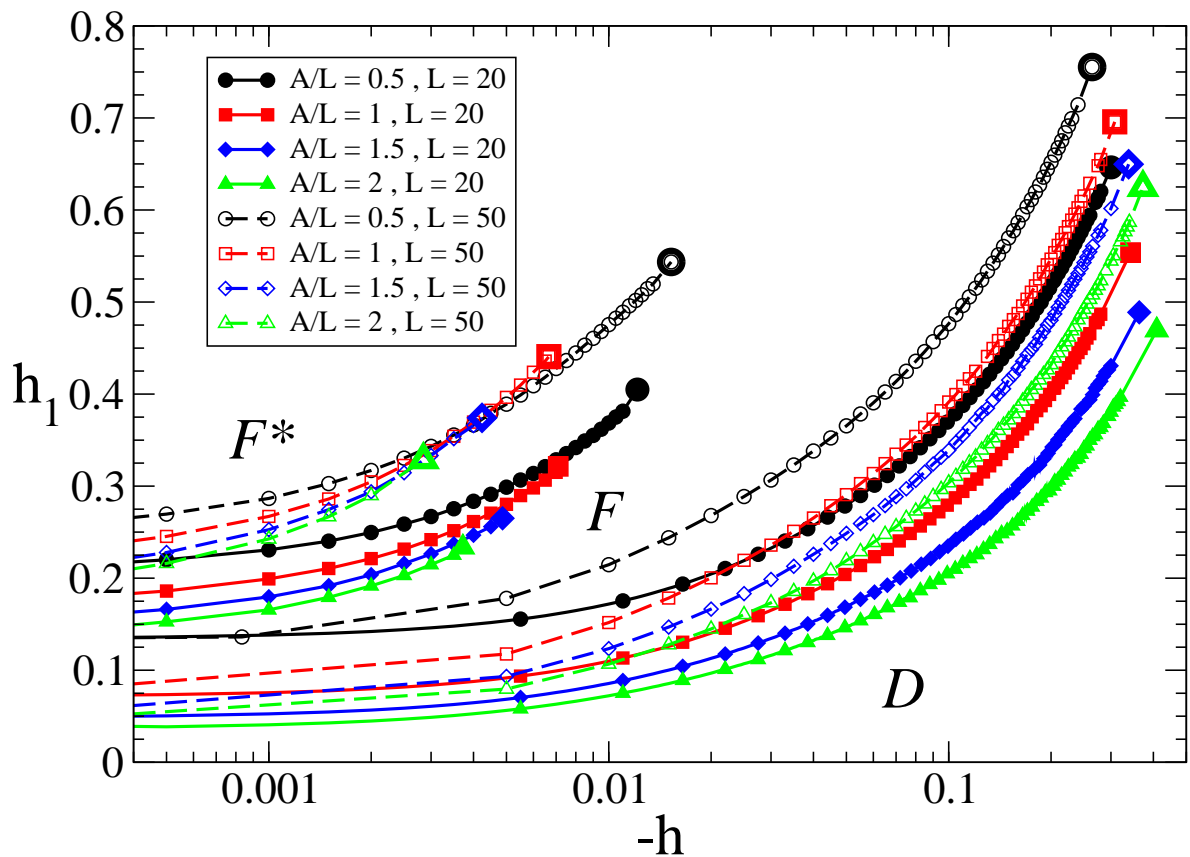}}
\caption{Off-coexistence $c=0$ interfacial phase diagram. The phase boundaries between $D$ and $F$ (filling transition lines)
and between $F$ and $F^*$ (prewetting transition lines) are represented by continuous lines/filled symbols for $L=20$ and  
dashed lines/open symbols for $L=50$. Circles correspond to the phase boundaries for $A/L=0.5$, squares correspond to 
$A/L=1.0$, diamonds correspond to $A/L=1.5$ and triangles correspond to $A/L=2.0$. Big symbols indicate 
the position of the critical points.}
\label{fig16}
\end{figure}

In general, there are two transitions between these interfacial states: the filling transition between $D$ and $F$ states, 
and a transition between $F$ and $F^*$ states, which we will denote as prewetting transition, as its characteristics are
reminiscent to those of the prewetting transition on the flat substrate, with the thickness of the liquid layer on 
the top of substrate as the order parameter. These transitions are first-order,
and they are located at the crossing of the different free-energy branches for constant $h$, analogously to the 
procedure followed for $h=0$. Figure \ref{fig14} shows the typical 
magnetization profiles of the coexisting states at filling and prewetting transitions, where $a$ is a $D$ state, $b$ and $c$
are $F$ states and $d$ is a $F^*$ state. Both filling and prewetting transition lines end up at critical points. Prewetting
is restricted to a small range of values of $h$ (as the prewetting line for the flat substrate). On the contrary,
the filling transition is observed for a wider range of $h$. For large enough values of $A$, 
the filling and prewetting transitions emerge from the filling and wetting transition points at bulk coexistence, i.e. 
$h=0$. However, if $A$ is small, both filling and prewetting transitions can exist even when at bulk coexistence there 
is no filling transition (i.e. in the Wenzel regime). Figure \ref{fig15} shows the interfacial phase diagram for $L=5$ 
and $A/L=0.5$. At bulk coexistence, there is only a wetting transition between a $D$ and a $W$ state at a value of $h_1$ 
close to the predicted by Wenzel law. For $h<0$ but close to bulk coexistence, a prewetting line where $D$ and $F^*$ 
states coexist emerges tangentially to the $h=0$ axis from Wenzel wetting transition, as expected from the 
Clausius-Clapeyron relationship. As the magnetization at the surface for the $D$ states is
negative, the midpoint interfacial height is taken as zero. By decreasing $h$, we reach to a triple point at $h_1\approx 0.153$ and $h\approx -0.0018$, where a $D$, $F$ and $F^*$ states coexist, and a filling and a prewetting transition lines 
emerge from this triple point.    
Both transitions end up at critical points, which are located by using the same technique as explained for the filling 
critical point in the $c=\infty$ case. Note that the midpoint interfacial height of the $F^*$ state in both
prewetting lines decreases as $-\log(-h)$, analogously to the thick layer phase along the prewetting of flat substrates. 
On the other hand, the midpoint interfacial height of the $D$ state along the filling transition remains zero until close
to the critical point. If $A$ is further decreased, both filling and $F-F^*$ prewetting transitions will eventually 
disappear, remaining only the $D-F^*$ prewetting transition. On the contrary, if $A$ is increased, the $D-F-F^*$ triple
point will be shifted towards $h=0$, and beyond this value the filling and prewetting transitions will become independent. 
Figure \ref{fig16} shows the off-coexistence phase diagram for $L=20$ and $L=50$ corresponding to different values
of $A/L$. As $h\to 0$, the filling and prewetting
lines tend to the states corresponding to the bulk coexistence filling and wetting transitions, respectively. 
For a given value of $L$, we observe that both filling and prewetting transitions shift towards lower values of 
$h_1$ as $A/L$ increases. The value of $|h|$ for the filling critical point increases as the substrate becomes rougher. 
On the contrary, the value of $|h|$ for the prewetting critical point decreases as $A/L$ increases. 
Regarding the dependence on $L$, both filling and prewetting lines shift towards higher values of
$h_1$ as $L$ increases for a given value of $A/L$. The range of values of $h$ where we observe the filling transition
line is reduced as $L$ increases, whereas for prewetting we observe different situations as $L$ is increased: for 
small $A/L$ the prewetting line range of values of $h$ increases, but decreases for larger values of $A/L$. It is 
interesting to note that, for $L=50$, the prewetting lines for different roughnesses seem to collapse in a master
curve for large $|h|$.

\begin{figure}[t]
\centerline{\includegraphics[width=\columnwidth]{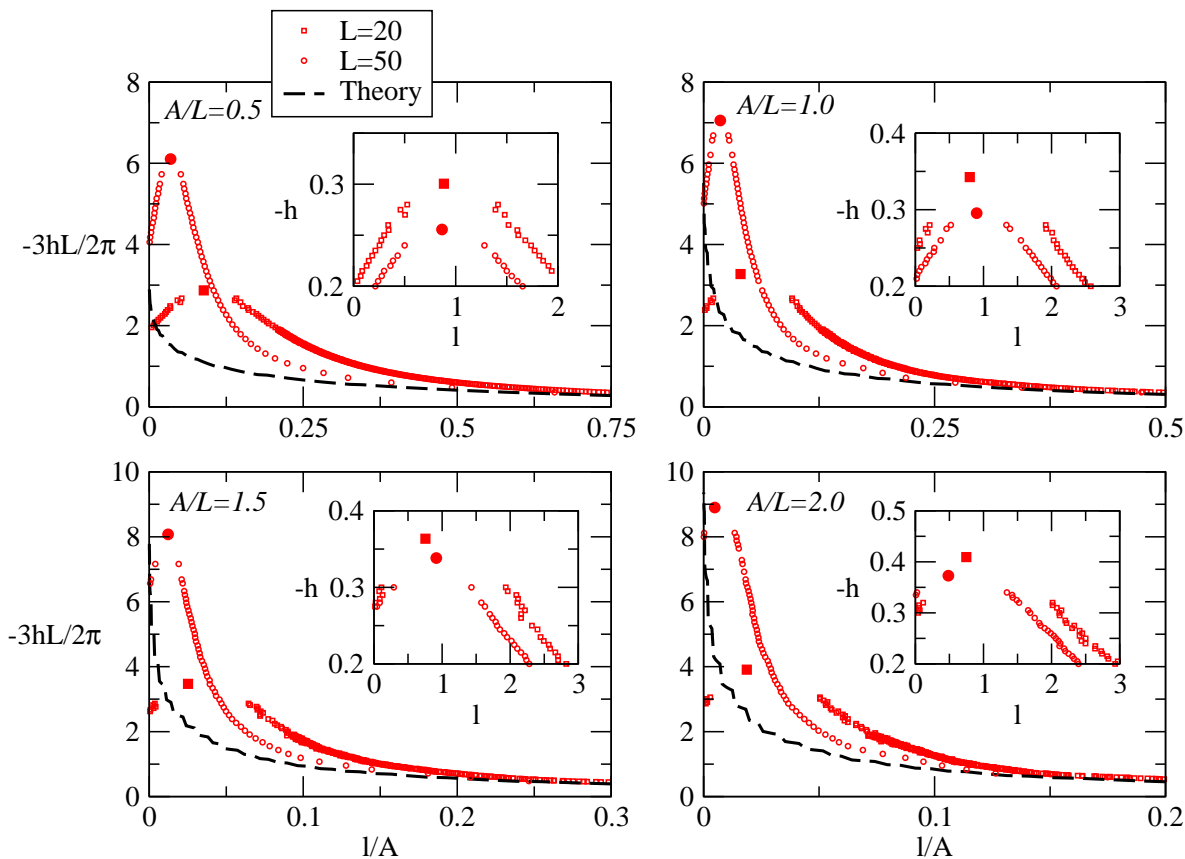}}
\caption{Plot of the midpoint interfacial height, in units of $A$, of the $D$ and $F$ states at the filling transition, as a 
function of $-3hL/2\pi$, corresponding to $c=0$ and $A/L=0.5,1,1.5$ and $2$, for $L=20$ (squares) and $L=50$ (circles). 
The dashed line corresponds to the macroscopic theory prediction for the midpoint interfacial height of the
$F$ state. The insets show a zoom of the $D$ and $F$ states midpoint interfacial heights, in units of the correlation 
length, as a function of $|h|$, close to the filling critical points (the meaning of the symbols is the same
as in the main plots).}
\label{fig17}
\end{figure}

A comparison with the macroscopic theory shows that, along the filling transition line, two different regimes can be 
observed. As explained in section \ref{sec2}, the macroscopic theory predicts that the filling transition also exists for
off-coexistence conditions. Within this theory, the contact angle $\theta$ at the filling transition
and $l/A$, where $l$ is the midpoint interfacial height of the coexisting $F$ state defined as (\ref{outcoexistence3}), 
are only functions of $qA$ and $qR$. We recall that $\theta$ is only function of $h_1$. 
On the other hand, by the Young-Laplace equation (\ref{laplace}) adapted to the Ising model, 
$R=\sigma_{+-}/(2m_0|h|)=1/(3|h|)$ in our units. So, the macroscopic theory predicts that, for a given value of $A/L$, both 
$h_1$ at the filling transition and $l/A$ are functions of $qR=2\pi/(3|h|L)$. However, this scaling is only obeyed for small
values of $|h|$. Figure \ref{fig17} shows the midpoint interfacial height of the $F$ states along the filling transition
line. We see that, for both $L=20$ and $L=50$ and all values of $A/L$, 
our numerical results coincide with the macroscopic theory prediction for small $h$ or, equivalently, large $l/A$.
However, as $l$ decreases, we see that the curve deviates from the theoretical prediction until reaches the critical
point of the filling transition. This deviation starts in all cases when $l\sim 5-10$, i.e. when the midpoint interfacial
height is of order of the correlation length, which occurs when $|h|\sim 1/L$. A closer insight on the magnetization
profiles show that the midpoint interfacial heights of both $D$ and $F$ states near the filling transition 
critical point behave similarly to the interfacial heights along the prewetting line of a flat substrate (compare insets 
of figure \ref{fig17} and figure \ref{fig23}). In fact, the filling transition line seems to converge to the 
prewetting line for the flat substrate as $L$ increases or $A$ decreases. So, there is a crossover from a geometrically
dominated behaviour at the filling transition for $|h|\lesssim 1/L$, to a prewetting-like behaviour for larger values of 
$|h|$, with can be regarded as a perturbation of the prewetting line with corrections due to the substrate curvature
at the bottom.   

\begin{figure}[t]
\centerline{\includegraphics[width=\columnwidth]{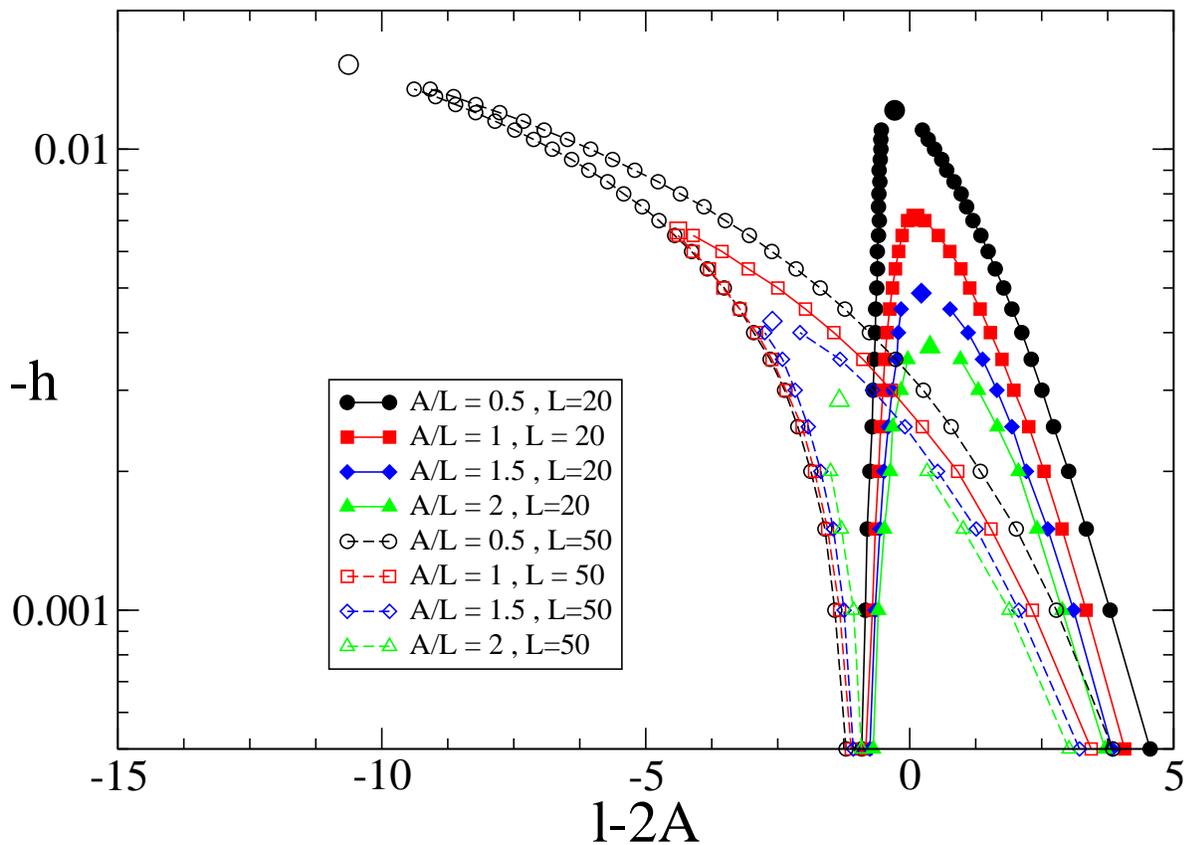}}
\caption{Plot of the midpoint interfacial height of the $F$ and $F^*$ states at the prewetting transition, as a
function of $h$, corresponding to $c=0$ and $A/L=0.5$ (circles), $A/L=1.0$ (squares), $A/L=1.5$ (diamonds) and 
$A/L=2$ (triangles), and $L=20$ (full symbols) and $L=50$ (open symbols). The largest symbols correspond to the locations
of the prewetting critical points.}
\label{fig18}
\end{figure}

Regarding the prewetting line, we observe that there is a strong $L$-dependence.
For large $L$, we expect that prewetting lines converge to the corresponding to the flat substrate. However, this 
convergence is very slow, as it occurs for the associated wetting transition. 
Figure \ref{fig18} shows the midpoint-interfacial height corresponding to the $F$ and $F^*$ states along the prewetting 
line. These coexistence curves show a high asymmetry associated to the interfacial curvature at $x=0$, which increases with
$L$ for a given substrate roughness. Alternatively, we may use the surface magnetization at $x=L/2$ as the order parameter 
for the prewetting transition. The prewetting coexistence dome is more symmetric, but the location of the critical points 
is virtually indistinguisible from the obtained by considering the midpoint interfacial height. However, we cannot use
the interfacial height above the substrate top at $x=L/2$ as order parameter, since the surface magnetization 
corresponding to the $F$ state is always negative. This fact is another indication of the slow convergence to the planar 
case.    

\subsection{Results for $c=\infty$, $h<0$}

\begin{figure}[t]
\centerline{\includegraphics[width=\columnwidth]{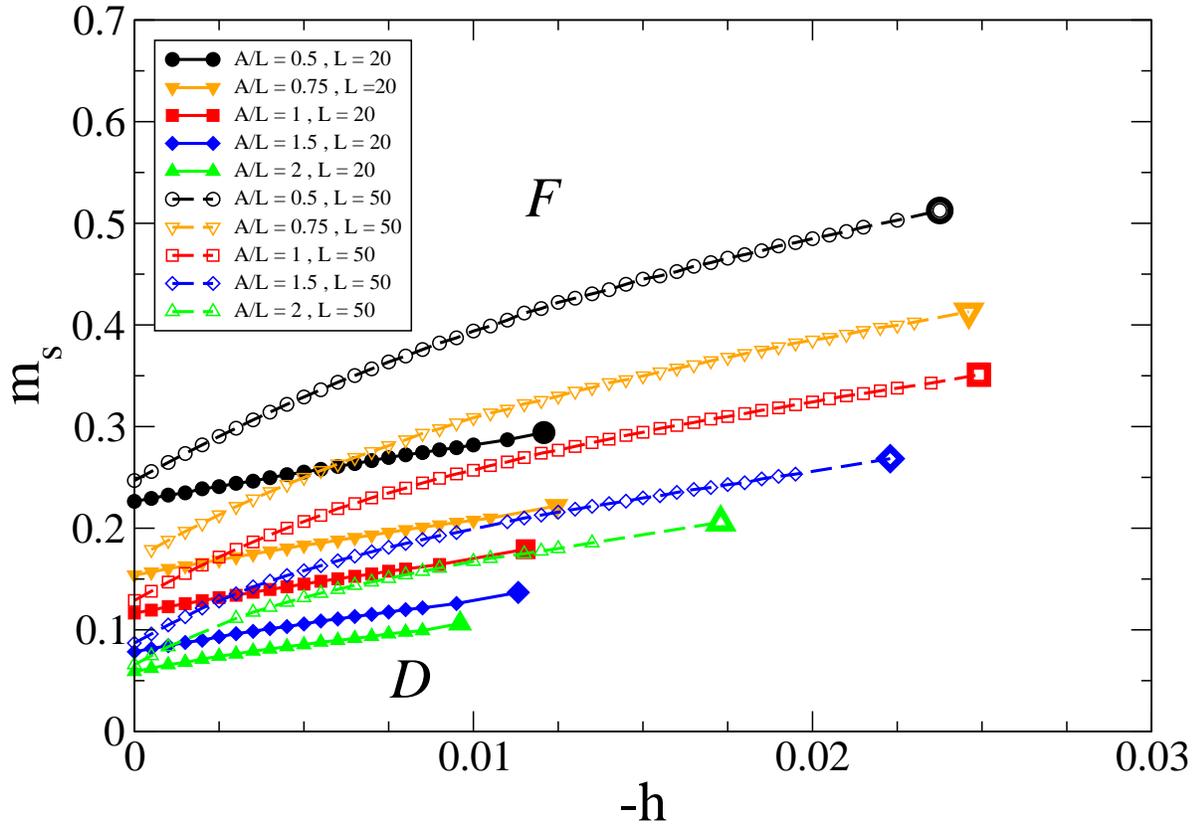}}
\caption{Off-coexistence $c=\infty$ interfacial phase diagram for $L=20$ (continuous lines/filled symbols) and $L=50$ 
(dashed lines/open symbols), corresponding to $A/L=0.5$ (circles), $A/L=0.75$ (triangles down), $A/L=1.0$ (squares), 
$A/L=1.5$ (diamonds) and $A/L=2.0$ (triangles up). The big symbols indicate the position of the critical points.}
\label{fig19}
\end{figure}

Finally, we turn back to the off-coexistence phase diagram for $c=\infty$. As in this case the wetting transition is always
continuous, only $D$ and $F$ states are observed for $h<0$, with characteristics similar to the corresponding states for
$c=0$. Thus we need only to focus on the filling transition between $D$ and $F$ states. If this transition exists at bulk 
coexistence, it has an off-coexistence extension which ends at a critical point. Thus, for a given value of $L$, the
filling transition line only exists for values of $A/L$ larger than the value of the roughness for which the critical 
filling occurs at bulk coexistence. Figure \ref{fig19} shows the 
off-coexistence phase diagram for $L=20$ and $L=50$, corresponding to different values of $A/L$. 
For each value of $L$ we observe that the range of values of $|h|$ of the filling transition line, which is given by
the value of $h$ for its critical point, is a non-monotonous function of the roughness: the critical value of $|h|$ 
for small values of $A/L$ increases, but it decreases for rougher substrates. This is in contrast with the interfacial
Hamiltonian model prediction which states that the value of $|h|$ at the critical point of the filling transition is
an increasing function of $A$ \cite{Rascon}. However, it captures correctly the observed feature that the filling 
transition is shifted towards lower values of $m_s$. When comparing distinct values of $L$ we see clear differences.
First of all, the filling transition line is almost linear for $L=20$ but it has some curvature for $L=50$. For a given
value of the roughness parameter $A/L$, the range of values of $m_s$ and $|h|$ for the filling transition line
increases with $L$. Although the filling transition lines start approximately at the same value (recall that there 
is some $L$-dependence on the filling transition at bulk coexistence), the slope of these lines for $h=0$ depends strongly 
on $L$. In fact, this dependence can be rationalized by the macroscopic theory, which predicts that $m_s$ along the
filling transition is a function of $2\pi/(3|h|L)$, as it was discussed for the $c=0$ case. As in the latter, we
observe a qualitative agreement with the macroscopic theory prediction only for small values of $|h|$. On the other hand,
the effective Hamiltonian model scaling behaviour of the off-coexistence filling transition \cite{Rascon}, which states that
for a given $A$ and regardless the value of $L$, $\Delta\tilde{T}=\sqrt{2}L(1-m_s)/(2\pi)$ is a function of 
$3|h|L^2/(2\pi)^2$, is completely broken down for our range of values of $A$. 

\begin{figure}[b]
\centerline{\includegraphics[width=\columnwidth]{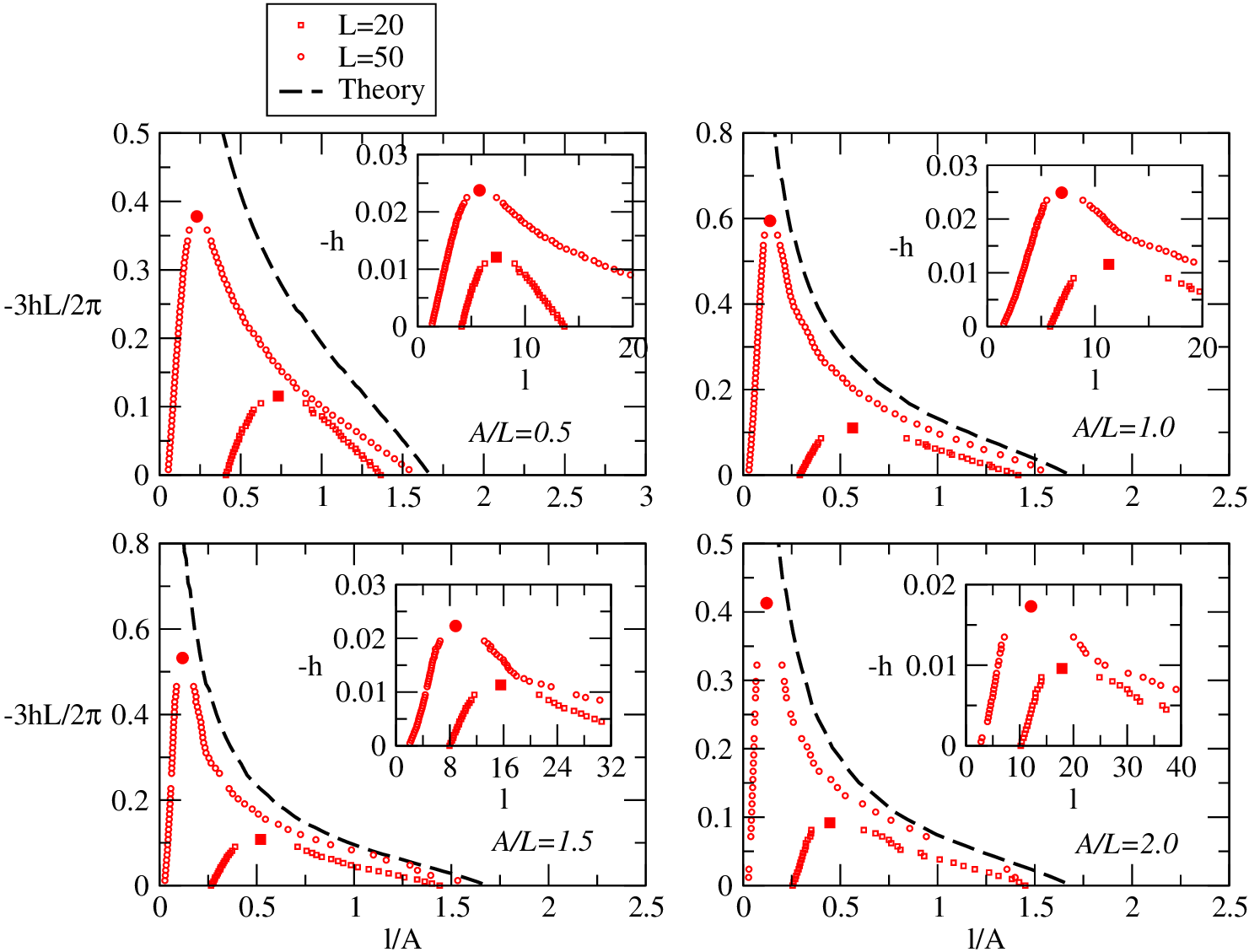}}
\caption{Plot of the midpoint interfacial height, in units of $A$, of the $D$ and $F$ states at the filling transition, as a 
function of $-3hL/2\pi$, corresponding to $c=\infty$ and $A/L=0.5,1,1.5$ and $2$, for $L=20$ (squares) and $L=50$ (circles). 
The dashed line corresponds to the macroscopic theory prediction for the midpoint interfacial height of the
$F$ state. The insets show a zoom of the $D$ and $F$ states midpoint interfacial heights, in units of the correlation 
length, as a function of $|h|$, close to the filling critical points (the meaning of the symbols is the same
as in the main plots).}
\label{fig20}
\end{figure}

Figure \ref{fig20} shows the midpoint interfacial heights of the $D$ and $F$ states along the off-coexistence filling 
transition. Note that, as $m_s$ is always positive, the $D$ state has a positive midpoint interfacial height for all values
of $h$. We see that the filling transition critical points have a midpoint interfacial height much larger than 
the bulk correlation length, although it decreases for larger $L$. This observation indicates that the emergence 
of the filling transition critical point for $c=\infty$ differs from the $c=0$ case. Unlike the $c=0$ situation, 
our numerical results show large deviations with respect to the macroscopic theory prediction for the midpoint interfacial
height of the $F$ states. Note that the macroscopic theory always overestimate the interfacial height, even for small $|h|$. However, as $L$ increases,
our numerical values seem to converge to the values obtained from the macroscopic theory. This suggests that for larger 
values of $L$ the macroscopic theory and the numerical results may agree, at least if the midpoint interfacial height 
remains much larger than the correlation length. However, the uncertainties introduced by our numerical method for 
larger values of $L$ prevented us to further explore this possibility.
    
\section{Discussion and conclusions \label{sec5}}

In this paper we study the fluid adsorption on sinusoidal substrates within the mean-field approach by using 
the Landau-Ginzburg model. We consider intermediate and rough substrates, i. e. values of the roughness parameter
$A/L\ge 0.5$. We focus on the filling, wetting and related phenomena under saturation conditions and off-coexistence, 
and compare our numerical results with approximate theories such as the macroscopic theory and effective interfacial 
models. Different scenarios are observed depending on the order of wetting transition for a flat substrate and the 
substrate period $L$. For small $L$ (i.e. $A$ of order of the correlation length) there is only a wetting transition
between $D$ and a $W$ interfacial states, of the same order as the wetting transition for the flat substrate. If 
first-order, it follows the phenomenological Wenzel law and has associated an off-coexistence prewetting line, 
while if critical it occurs at the same temperature as for the flat substrate. On the other hand, for large $L$ the 
interfacial unbinding occurs via two steps: a filling transition between $D$ and $F$ states, and a wetting transition
between $F$ and $W$ states. The filling transition is always first-order, occurs under the conditions predicted by
the macroscopic theory (although the agreement is quantitatively better for first-order wetting substrates) and 
has an off-coexistence extension which ends up at a critical point. Wetting is of the same order as for the flat case.
If first-order, it shows a significant shift towards lower temperatures with respect to the wetting temperature
for the flat substrate, while for critical wetting it occurs precisely at the same temperature. These features are not 
explained by the macroscopic theory and are in agreement with interfacial Hamiltonian theories. Note that wetting always
occurs when the liquid-vapour interface of the $F$ state is near the substrate top, so we argue that it is controlled
by the geometric characteristics and wetting properties of the substrate around the top. The borderline between the
small and large-$L$ scenarios is also dependent on the order of the wetting transition of the flat substrate. So, the filling
transition disappears at a $D$-$F$-$W$ triple point for first-order wetting, while for critical wetting the filling
transition line ends up at a critical point, as predicted by interfacial models. Finally, regarding the off-coexistence
transitions, the filling transition line agrees with the macroscopic theory predictions if $L$ is large enough and very
close to bulk coexistence, i.e. $|h|\lesssim 1/L$. Under these conditions, the midpoint interfacial height of the 
$F$ state is much larger than the bulk correlation length. Again, the agreement with the macroscopic theory worsens for
critical wetting substrates. For first-order wetting substrates, we observe a crossover
to a prewetting-like behaviour for larger values of $|h|$. This line is different from the proper prewetting line, which
is associated to the wetting properties at the substrate top.    

Although we have considered only two extreme situations ($c=0$ for first-order wetting substrates and $c=\infty$ for
critical wetting substrates) we expect similar scenarios for small or large $c$, respectively. 
The borderline between these scenarios
is expected to occur around the tricritical wetting conditions for the flat substrate. This study is beyond the present
work. Our model overcomes many of the problems with previous approaches, such as the neglected role of intermolecular 
forces in the macroscopic theory, or the appropriate form of the binding potential for the effective interfacial models 
for rough substrates \cite{14,15}. However, the simplicity of our functional have additional disadvantanges. For example, it 
does not describe properly the packing effects close to the substrates due to the hard-core part of the intermolecular
interactions, which is of order of the bulk correlation length away from the bulk critical point. As a consequence, 
the phenomenology for small $L$ may be affected by these effects, and even for larger values of $L$ some of the predicted
transitions may be preempted by surface or bulk solidification. On the other hand, our functional is appropriate
for short-ranged intermolecular forces, although in nature dispersion forces are ubiquous. In order to take into account 
the packing effects or long-ranged interactions, more accurate functionals should be used. 

Finally, in our study interfacial fluctuations are completely neglected due to its mean-field character. These may have
an effect for the continuous transition. For short-ranged forces, $d=3$ is the upper critical dimension for critical 
wetting of a flat substrate. So, we anticipate that capillary wave fluctuations may alter the critical behaviour of 
the critical wetting on the flat substrate. Furthermore, interfacial fluctuations may have more dramatic effects
in transitions such as filling. In fact, although filling is effectively a two-dimensional transition (as it is prewetting),
the interfacial fluctuations are highly anisotropic, since the interfacial correlations along the grooves axis are much 
stronger than across different grooves. This may lead to a rounding of the filling transition due to its
quasi-one dimensional character, as it happens for single grooves. Further work is required
to elucidate the effect of the interfacial fluctuations in the adsorption of rough substrates.
\ack

We thank Dr. P. Patr\'{\i}cio, Dr. N. R. Bernardino, Prof. A. O. Parry and Dr. C. Rasc\'on for illuminating discussions. 
A. R.-R. and J. M. R.-E. acknowledge financial support from the Spanish Ministerio de Econom\'{\i}a y Competitividad through
grants no. FIS2009-09326 and FIS2012-32455, and Junta de Andaluc\'{\i}a through grant no. P09-FQM-4938, all co-funded
by the EU FEDER, and the Portuguese Foundation for Science and Technology under Contract No. EXCL/FIS-NAN/0083/2012.

\appendix
\section{The Landau-Ginzburg theory of wetting of flat substrates\label{appendix}}

In this section we review the Landau-Ginzburg theory, focusing on its application to interfacial transitions such as
wetting transition. Because of its simplicity, this model has been extensively studied in this context in the 
literature \cite{Cahn,Nakanishi1,Nakanishi2,Pandit,Pandit2,Brezin,Margarida,Jin1,Jin3,Jin2,Jin4}.
For convenience we use the magnetic language, where the order parameter has the same symmetry as the magnetization per 
unit volume in the Ising model. However, the results obtained for this system are completely valid for the interfacial 
phenomenology of simple fluids, identifying the order parameter with the deviation of the density with respect to its 
critical value. We will also restrict ourselves to the three-dimensional situation, although the formalism can be applied 
to other dimensionalities.
 
The free energy functional of the system in contact with a substrate can be expressed in terms of the order parameter 
field $m(\mathbf{r})$ as:
\begin{eqnarray}
{\cal F}&=&{\cal F}_0+\int_{V} d\mathbf{r}\left[\frac{g}{2}(\bnabla m)^{2}+a_{2}tm^{2}+a_{4}m^{4}-hm\right]\nonumber\\
&+&\int_{\cal S} d\mathbf{s} \frac{c}{2}(m(\mathbf{s})-m_{s}(\mathbf{s}))^2
\label{free-energy}
\end{eqnarray}
where the first term corresponds to Landau-Ginzburg functional on total volume $V$ while the second term takes into 
account the interaction with the substrate. Thus, ${\cal F}_0$ is a reference free energy, $g$, $a_2$ and $a_4$ 
are positive constants, $h$ is the ordering field (magnetic field magnetic systems, deviation of chemical potential 
with respect to the value at coexistence in fluid systems) and $t=(T-T_c)/T_c$ characterizes
the temperature deviation with respect to the critical value $T_c$. Regarding the interaction with the substrate, 
the integration is restricted to the surface of the substrate $S$. Finally $c$ is the \emph{enhancement} parameter
and $m_{s}(\mathbf{s})$ is the favoured order parameter value by the substrate, and that will be assumed to be positive, 
so it favors the phase with volume order parameter  $+m_0$ when $t<0$ and $h=0$. For later purposes, it will be useful 
to define the applied surface field $h_1(\mathbf{s})={cm_{s}}(\mathbf{s})$.
For theoretical analysis and its computational implementation it is convenient to use a description in terms of 
reduced units. To do this, we must first determine the natural scales of each
variables. The natural scale for the order parameter field is
given by the equilibrium value of this magnitude $m_{eq}$ for $h=0$ in the Landau theory.
Although for $t>0$ $m_{eq}= 0$, for temperatures below the critical
and $h=0$ the states characterized by $m_{eq}=\sqrt{a_{2}(-t)/(2a_{4})}$
and $-\sqrt{a_{2}(-t)/(2a_{4})} $ are at coexistence. Thus, as we are
interested in situations where there is coexistence of phases (which implies
that $t<0$), we define $m_0$ as:
\begin{equation}
m_0=\sqrt{\frac{a_{2}|t|}{2a_{4}}}
\end{equation}
On the other hand, the natural length scale is given by the correlation length $\xi$ defined from the Ornstein-Zernike 
theory correlation applied to the Landau-Ginzburg functional:
\begin{eqnarray}
\xi&=&\sqrt{\frac{g}{2 a_2 t + 12 a_4 m_{eq}^2}}\nonumber\\&=&
\cases{
\sqrt{\frac{g}{2a_2 t}} &$t>0$ \\
\sqrt{\frac{g}{4 a_2 (-t)}} &$t<0$ \\
}
\end{eqnarray}
By analogy with the definition of the scale of the order parameter, define the length scale $\xi_0$ as
\begin{equation}
\xi_0=\sqrt{\frac{g}{4 a_2 |t|}}
\label{xi0}
\end{equation}
Therefore, if we define $\tilde m=m/m_0$ and $\tilde \mathbf{r}=\mathbf{r}/\xi_0$
Then we can define a reduced free energy as:
\begin{eqnarray} 
\tilde {\cal F}&=&\frac{{\cal F}}{8a_{4}m_{0}^{4}\xi^{3}_0}=\tilde {\cal F}_0\nonumber\\&+&\int_{\tilde V}
d{\tilde \mathbf{r}}
\left[\frac{1}{2}(\tilde{\bnabla} \tilde{m})^{2}-\tilde h \tilde m 
\pm\frac{1}{4}\tilde
m^{2}+\frac{1}{8}\tilde{m}^4\right]\nonumber\\&+&\int_{\tilde {\cal S}} d\tilde \mathbf{s}
\frac{\tilde c}{2}(\tilde m-\tilde m_{s})^2 
\label{redfree-energy}
\end{eqnarray}
where $\tilde {\cal F}_0={\cal F}_0/(8a m_{0}^{4}\xi^{3}_0)$, $\tilde{\bnabla}=\xi_0\bnabla$, and positive or negative
corresponds to $t>0$ or $t<0$, respectively. The reduced ordering field $\tilde h$ is defined as:
\begin{equation}
\tilde h=\frac{h}{8a_{4}m_{0}^{3}}
\end{equation}
Finally, the reduced parameters of the interaction with the surface are defined as $\tilde m_ {s}= m_{s}/m_0$ and 
$\tilde c=c/(8a_{4}m_{0}^{2}\xi_0)$. Consequently, the reduced surface field ${\tilde h_1}(\tilde\mathbf{s}) = 
h_1 (\mathbf{s})/(8a_{4} m_{0}^{3} \xi_0)$. Since there is some freedom to choose the source of energy,
choose $\tilde F_0=\tilde V/8$. Thus, we can rewrite (\ref{redfree-energy}) as:
\begin{eqnarray}
\tilde F&=&\int_{\tilde V} d{\tilde\mathbf{r}} \left[\frac{1}{2}(\tilde{\bnabla}\tilde{m})^{2}-\tilde h \tilde m
+\frac{1}{8}(\tilde{m}^2\pm 1)^2\right]\nonumber\\&+&\int_{\tilde {\cal S}} d\tilde\mathbf{s}
\frac{\tilde c}{2}(\tilde m-\tilde m_{s})^2
\label{redfree-energy2}
\end{eqnarray}
Hereafter we will only consider reduced units, so we will drop the tildes in the expressions above. 
In the mean field approximation, the equilibrium profile parameter order is obtained by minimization 
of the functional (\ref{redfree-energy2}) \cite{Brezin}.
Using the functional derivative of $F$ with respect to
$m(\mathbf{r} _0)$ (assuming $ \mathbf{r} _0 $ is not on the substrate) and
making it equal to zero, we obtain the following Euler-Lagrange equation:
\begin{equation}
\nabla^2 m= -h + \frac{m(m^2\pm 1)}{2}
\label{EL}
\end{equation}
On the other hand, the variation of the order parameter field in a surface point $\mathbf{s}_0$ leads to the following 
boundary condition:
\begin{equation}
\mathbf{n \cdot} \bnabla m (\mathbf{s}_0) = c(m(\mathbf{s}_0)-m_{s}(\mathbf{s}_0))
=cm(\mathbf{s}_0)-h_1(\mathbf{s}_0)
\label{cc1}
\end{equation}
where $\mathbf{n}$ is the inward normal to the substrate in $\mathbf{s}_0$ (i.e. directed towards
the system). Finally, we impose that the order parameter far from the surface takes
the equilibrium value given by the Landau theory $m_b$:
\begin{equation}
m(\mathbf{r})\to m_b \qquad\mbox{far from the substrate}
\label{cc2}
\end{equation}
If we restrict ourselves to the situation of coexistence ($h=0$ and $t<0$),
the order parameter far from the substrate has the boundary condition $m\to -1$.

In general, the differential equation (\ref{EL}) with boundary conditions (\ref{cc1}) and (\ref{cc2}) cannot be
solved analytically and we must resort to numerical methods. However, in the case of a flat substrate with $c$ and $h_1$ 
constants the problem can be solved analytically. Consider that the substrate is on the plane $xy$. Then, by symmetry, 
the order parameter field depends only of the coordinate $z$. At bulk coexistence ($h=0$ and $t<0$),
(\ref{EL}) reduces to:
\begin{equation}
\frac{d^2 m}{dz^2}=\frac{m(m^2-1)}{2}
\label{EL2}
\end{equation}
Multiplying this equation by $(dm/dz)$ and integrating in the range $z \in [z_0, +\infty]$, we obtain the following 
expression:
\begin{equation}
\frac{1}{2} \left(\frac{dm}{dz}\right)^2\Bigg |_{z=z_0}=\frac{1}{8}(m(z_0)^2-1)^2
\label{1stintegral}
\end{equation}
where we have used the boundary condition (\ref{cc2})
($m(z\to\infty) \to -1)$ and that $dm/dz \to 0 $ when $ z \to \infty $.
The order parameter profile thus will be a monotonous increasing function if $m(0) <-1 $, and a decreasing function 
otherwise.  We can obtain from (\ref{1stintegral}) the derivative of
order parameter to an arbitrary height $z$ as:
\begin{equation}
\frac{dm}{dz}=-\frac{1}{2}(m+1)|m-1| 
\label{1stintegral2}
\end{equation}
This condition is valid for all $ z \ge 0 $. In fact, we can integrate
(\ref{1stintegral2}) for $m(z)$. So, for $ m(0)< 1$
the equilibrium profile satisfies order parameter
\begin{equation}
m(z)=-\tanh\left(\frac{z-z_0}{2}\right)
\label{perfil1}
\end{equation}
where $z_0=2\textrm{ atanh } m (0) = \ln [(1 + m (0)) / (1-m (0 ))]$. If $ z_0> 0 $,
these profiles describe interfacial states where you can identify
a layer of bulk order parameter $+1$ for $z<z_0$
in contact with the bulk phase characterized by the order parameter
$-1$ for $ z> z_0 $. Therefore, the interfacial position is given by $z_0$ as $m(z_0)=0$.

\begin{figure}[t]
\centerline{\includegraphics[width=\columnwidth]{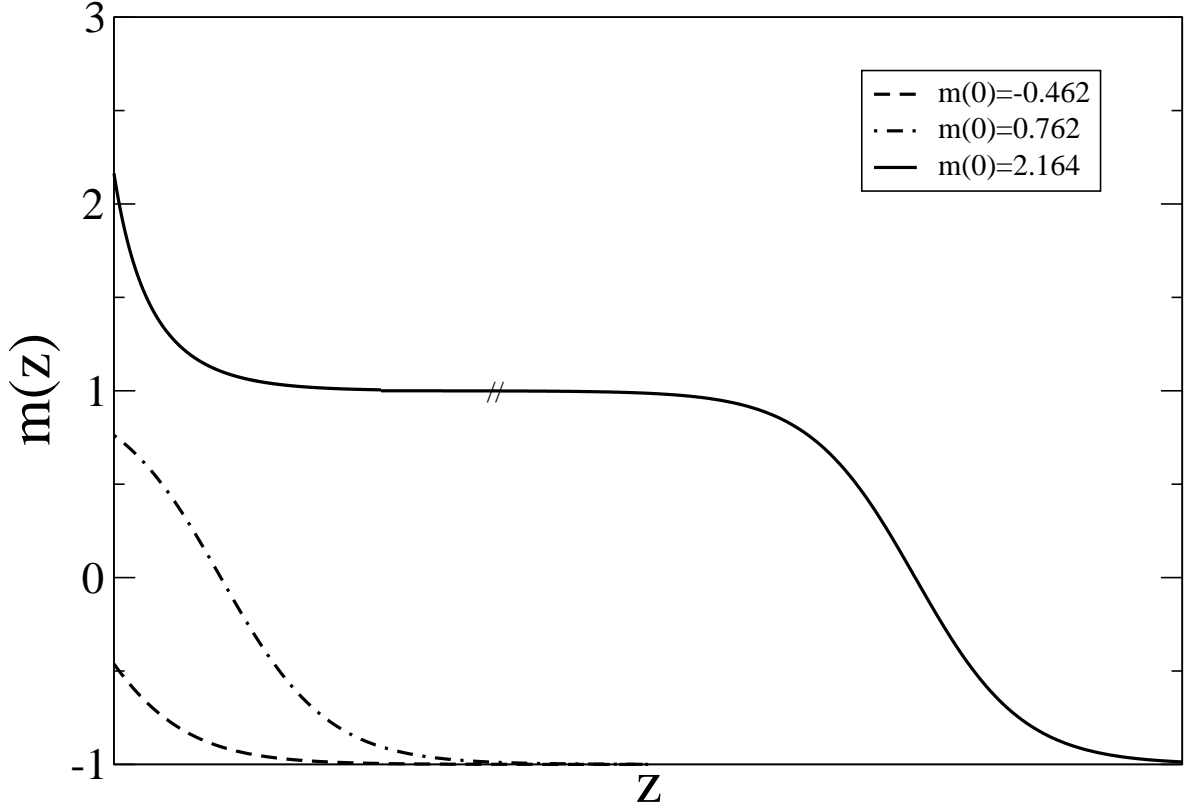}}
\caption
{Typical profiles of order parameter theory
Landau-Ginzburg for the phenomenon of wetting on a flat substrate. The
profiles corresponding to $m(0)=-0.462$ (dashed curve), $m(0)=0.762$
(curve of dots and dashes) and $m(0)=2.164$. The latter case corresponds to complete wetting, so
the width of the layer of $+1$ phase becomes infinite.
}
\label{fig21}
\end{figure}

If $ m(0)>1$, the solution has the expression:
\begin{equation}
m(z)=
\cases{
\coth\left(\frac{z-z_0}{2}\right) & if $m(z)>1$\\
-\tanh\left(\frac{z-l}{2}\right) & if $m(z)<1$\\
}
\label{perfilwetting}
\end{equation} 
where $ z_0 =-2\textrm{ acoth } m (0) =-\ln [(m (0) + 1) / (m (0) -1)] $. However,
it is easy to see from the first equation that $m(z)>1 $ for all $z$.
This implies that the only allowable value of $ l $ is infinite. As at $z=l$,  
$m(l)=0$, this condition implies that a layer of 
infinite thickness of order parameter $+1$ has nucleated between the substrate and the bulk phase.
Therefore,  profiles with $m(0)<1$ will correspond to 
\emph{partial wetting}, while if $m(0)>1$ we have \emph{complete} wetting. Figure \ref{fig21} shows some typical order 
parameter profiles.
The equilibrium free energy $F_{eq}$ can be obtained by replacing the order parameter profiles in the functional
(\ref{redfree-energy2}). However, their evaluation can be simplified taking into account  
(\ref{1stintegral}). If we define the surface tension between the substrate and the bulk  phase with order parameter
$-1$, $\sigma_{W-} $, as $F_ {eq}/A $ (note that the bulk contribution is zero in our case), then we can evaluate it as:
\begin{equation}
\sigma_{W-}=\int_0^\infty \left(\frac{dm}{dz}\right)^2 dz+
\frac{c}{2}(m(0)-m_{s})^2
\label{sigmaw-1}
\end{equation}
Given that the order parameter profiles are monotonous, and
using (\ref {1stintegral2}), we can express the surface tension as:
\begin{eqnarray}
\sigma_{W-}=-\int_{-1}^{m(0)} \left(\frac{dm}{dz}\right) dm+
\frac{c}{2}(m(0)-m_{s})^2
\nonumber\\=\int_{-1}^{m(0)} \frac{1}{2}(m+1)|m-1|dm +
\frac{c}{2}(m(0)-m_{s})^2 
\label{sigmaw-2}
\end{eqnarray}
Therefore, the surface tension $ \sigma_{W-} $ has the expression:
\begin{equation}
\sigma_{W-}=
\cases{
\begin{array}{l} \frac{m(0)}{2}-\frac{m(0)^3}{6}+\frac{1}{3}\\+\frac{c}{2}(m(0)-m_{s})^2 
\end{array}& $m(0)<1$\\
\sigma_{W+}+\sigma_{+-}& $m(0)>1$}
\label{sigmaw-3}
\end{equation}
where $ \sigma_{+-}$ is the surface tension associated to the interface
between the two bulk phases at coexistence:
\begin{equation}
\sigma_{+-}=\int_{-1}^{1} \frac{1}{2}(1-m^2)dm = \frac{2}{3}
\label{sigma+-}
\end{equation}
and $\sigma_{W +}$ is the surface tension between the substrate and a bulk
phase with order parameter $+1$. Under these conditions, (\ref{1stintegral2}) changes to 
$dm/dz =-|m+1|(m-1)/2 $, and then
\begin{eqnarray}
\sigma_{W+}=-\int_{1}^{m(0)} \left(\frac{dm}{dz}\right) dm+
\frac{c}{2}(m(0)-m_{s})^2
\nonumber\\=\int_{1}^{m(0)} \frac{1}{2}|m+1|(m-1)dm +
\frac{c}{2}(m(0)-m_{s})^2
\nonumber\\
=-\frac{m(0)}{2}+\frac{m(0)^3}{6}+\frac{1}{3}+\frac{c}{2}(m(0)-m_{s})^2
\label{sigmaw+}
\end{eqnarray}
Note that the expression for $\sigma_ {W-}$ and $m(0)>1$ given by
(\ref{sigmaw-3}) is predicted by Young's law under conditions
of complete wetting.

\begin{figure}[t]
\centerline{\includegraphics[width=\columnwidth]{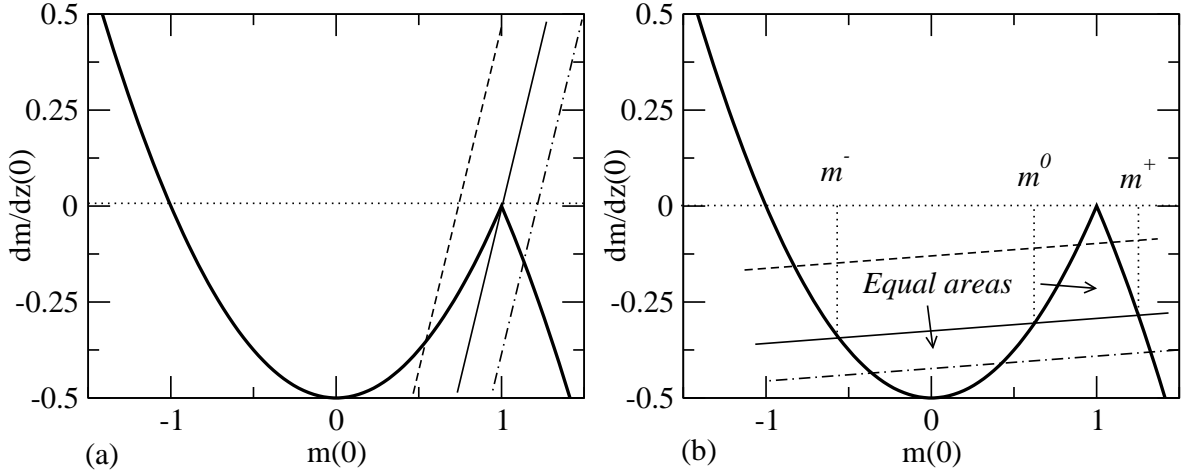}}
\caption
{Graphical construction to obtain $m(0)$ for: (a) $c>1$,
(b) $c<1$. The dashed lines correspond to partial wetting situations,
the dot-dashed lines to complete wetting situations, and the continuous lines
to the wetting transition.}
\label{fig22}
\end{figure}

\begin{figure}[t]
\centerline{\includegraphics[width=\columnwidth]{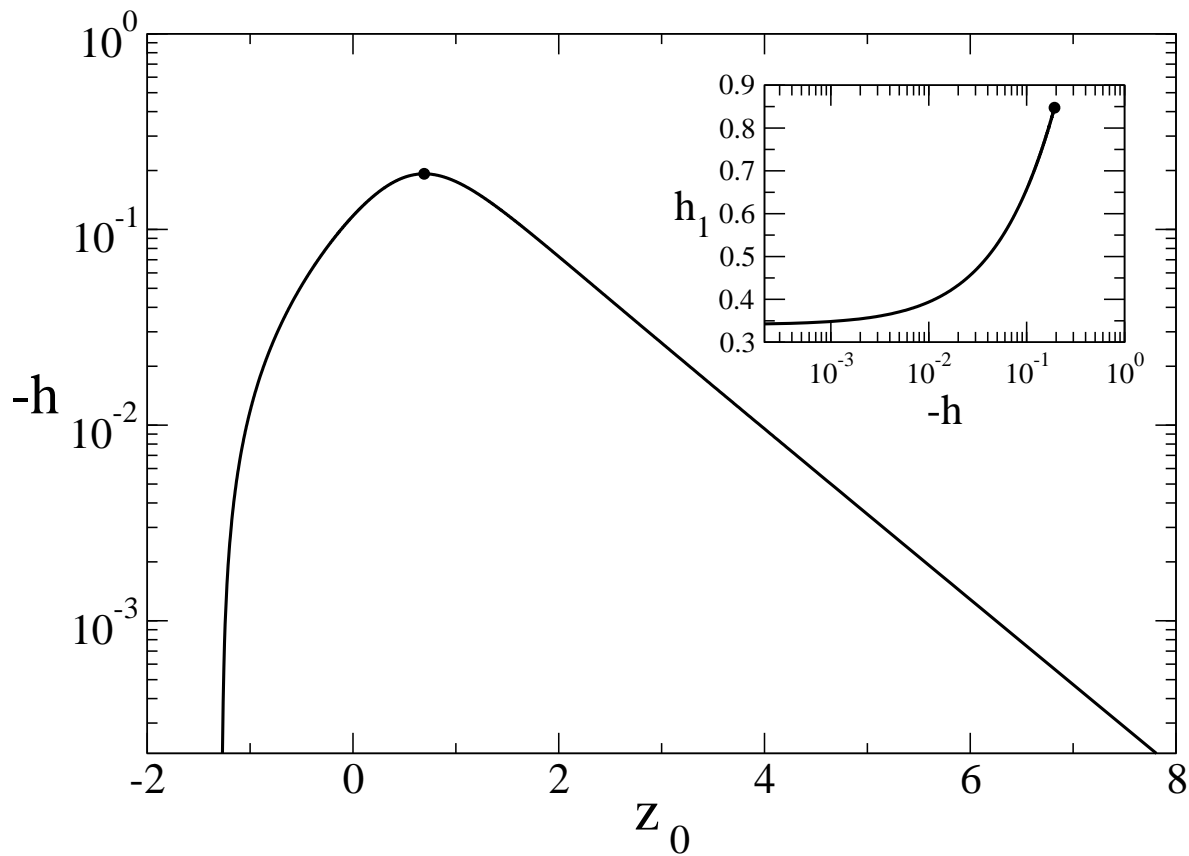}}
\caption
{Plot of the interfacial height $z_0$ of the coexisting interfacial states along the prewetting line for $c=0$ 
as a function of $|h|$. The dot corresponds to the location of the prewetting critical point, and negative values of 
$z_0$ means that the magnetization at the wall is negative. Inset: Plot of the prewetting line for $c=0$ on the $|h|\--h_1$ 
plane.}
\label{fig23}
\end{figure}

The values of $m(0)$ can be obtained by using the boundary condition
(\ref {cc1}). Therefore, the derivatives of the order parameter
profile at must satisfy simultaneously that:
\begin{equation}
\frac{dm}{dz}\Bigg|_{z=0}=-\frac{1}{2}(m(0)+1)|m(0)-1| 
=cm(0)-h_1
\label{graphical-solution}
\end{equation} 
Therefore, we get $m(0)$ by using a graphical construction (see figure \ref{fig22}).
Three different situations can be observed depending on the value of $ c $:
\begin {enumerate}
\item {\bf Critical wetting transition ($c>1$)}.
Under these conditions, there is only one intersection of (\ref{graphical-solution}) 
at a value of $m(0)$ given by the expression:
\begin{equation}
m(0)=
\cases{
c-\sqrt{c^2+(1-2cm_{s})}\quad  m_{s}<1\\
-c+\sqrt{c^2+(1+2cm_{s})}\quad m_{s}>1\\
}
\end{equation}
Therefore, the system goes continuously from a 
partial wetting situation for $ m_{s} <$ 1 to a situation
complete wet for $ m_{s}> 1$, so the
wetting transition is continuous and occurs at $m_{s}=1$.
\item {\bf Tricritical wetting transition ($c=1$)}. This situation corresponds to the borderline between 
continuous and first-order wetting transitions. However, the description of the transition is
similar to the case of critical wetting.
\item {\bf First-order wetting transition ($c<1$)}. Under these conditions, there may be up to three
intersections of (\ref{graphical-solution}), which will be 
denoted by $m^-$, $m^0 $ and $m^+$:
\begin{eqnarray}
m^-=c-\sqrt{c^2+(1-2h_1)} \label{m-}\\
m^0=c+\sqrt{c^2+(1-2h_1)} \label{msup0}\\
m^+=-c+\sqrt{c^2+(1+2h_1)} \label{m+} 
\end{eqnarray}
Therefore, there may be up to three possible order parameter profiles. To identify the true equilibrium profile, we
evaluate the surface free energy. It can be shown that the profile for the solution
term $m^0$ always has a free energy higher than other states. As for $h_1$ small or negative the only possible solution
corresponds to that with $m(0)=m^-$, and for large $h_1$ the only solution corresponds to the complete wetting profile 
where $m(0) = m^+ $, there must be an intermediate value
of $h_1$, where both states coexist. Thus, the wetting transition is given by the value of $h_1$ for which 
$\sigma_{W-} (m^-) = \sigma_{W+} (m^+)$. This condition has a graphical interpretation: the wetting transition occurs for
the value of $h_1$ for which the areas enclosed by curves
given by (\ref{graphical-solution}) between $m^-$ and $m^0$, on one
hand, and $m^0$ and $m^+$, on the other hand, are equal (Maxwell construction).
Starting from the wetting transition, and out of coexistence (i.e. $h<0$), the prewetting transition emerges \cite{Pandit}, 
where two distinct interfacial structures characterized by different but finite adsorbed phase film thicknesses (see 
figure \ref{fig23}). This transition line starts tangentially to the bulk coexistence curve $h=0$, as predicted by
the Clausius-Clapeyron relationship \cite{Hauge}, and finishes at the prewetting critical point.
The location of this transition is obtained by a similar construction to the outlined above for the wetting transition: 
the two interfacial phases are determined by the surface magnetization, obtained by the intersection of the following 
curves: 
\begin{eqnarray}
\frac{dm}{dz}\Bigg|_{z=0}&=&-\sqrt{-2h(m+1)+\frac{1}{4}(m^2(0)-1)^2}\nonumber\\ 
\frac{dm}{dz}\Bigg|_{z=0}&=&cm(0)-h_1
\label{graphical-solution2}
\end{eqnarray} 
Up to three solutions may be obtained. The true equilibrium profile corresponds to the solution with minimum surface free
energy. Coexistence is obtained when the areas enclosed by the curves given by (\ref{graphical-solution2}) are equal, and
the prewetting critical point corresponds to the situation where the three solutions merge into the same value.
\end{enumerate} 
Finally, we can obtain analytically the contact angle $\theta$ via Young's law:
\begin{equation}
\cos\theta=\frac{\sigma_{W-}(m(0))-\sigma_{W+}(m^*(0))}{\sigma_{+-}}
\label{costheta}
\end{equation}
where $\sigma_{W-}$, $\sigma_{W +}$ and $\sigma_ {+-}$ are given
by the expressions (\ref{sigmaw-3}) (\ref{sigmaw+}) and (\ref{sigma+-}),
respectively, $ m (0) $ is the value of the equilibrium order parameter
at $z=0$ (obtained through the construction explained above),
and $ m^*(0)$ is the value of the equilibrium order parameter at $z=0$ which 
decays to $m=1$ when $z \to +\infty$. This value can be obtained
via a graphical construction similar to that already explained,
where we look for solutions of the equation:
\begin{equation}
-\frac{1}{2}|m^*(0)+1|(m^*(0)-1)=cm^*(0)-h_1
\label{graphical-solution3}
\end{equation} 
For $h_1>0$, this solution is given by:
\begin{equation}
m^*(0)= 
-c+\sqrt{c^2+(1+2h_1)} \label{mstar} 
\end{equation}

In order to finish this introduction to the Landau-Ginzburg model of wetting of flat substrates, it is common in the 
literature to study the wetting phenomena by using interfacial Hamiltonians, where the surface free energy associated to
an interfacial configuration (i.e. by fixing the surface at which the magnetization is zero, for example), is given
by:
\begin{equation}
{\cal F}=\int_{\cal A} d\mathbf{s} \left[\frac{\sigma_{+-}}{2}(\bnabla\ell(\mathbf{s}))^2+W(\ell(\mathbf{s}))\right]
\label{interfacialhamiltonian}
\end{equation}
where $\ell(\mathbf{s})$ is the interfacial height above the position $\mathbf{s}$ of the substrate and $W(\ell)$
is the interfacial binding potential. A considerable 
work has been reported in the literature to justify (\ref{interfacialhamiltonian}) from first principles
\cite{Brezin,Jin1,Jin3,Jin2,Jin4}. More recently, a new derivation of (\ref{interfacialhamiltonian}) has been proposed, 
where in general the binding potential is not a local function but a non-local functional of $\{\ell(\mathbf{s})\}$
\cite{14,15,16,17,18,19,20}. For parallel and flat substrate and interface, this functional reduces again to a function, 
which has a long-distance expansion \cite{Brezin,Jin1,Jin2,15,20}:
\begin{eqnarray}
W(\ell)&\approx& -2h\ell -\frac{4c}{1+c}(1-m_{s})\exp(-\ell)\nonumber\\&+&\frac{4(c-1)}{1+c}\exp(-2\ell)+\ldots
\label{bindingpotential}
\end{eqnarray}
For critical wetting, this expansion is enough to characterize the divergence of the interfacial height. The equilibrium 
height is given by the absolute minimum of (\ref{bindingpotential}). So, at $h=0$, $\ell_{eq}=\ln(2(c-1)/(c(1-m_{s})))
\approx \ln(2/(1-m(0))) \approx \ln((1+m(0))/(1-m(0)))$ when $m_{s}\to 1$, in agreement with the full Landau-Ginzburg
model results.

\end{document}